\documentclass[twocolumn]{aa_old}
\usepackage{graphics}
\input{epsf}
\newcommand{\bfo}[1]{\mbox{\boldmath $#1$}}
\def\bvarphi{\mbox{\boldmath $\varphi$}}

\begin{document}
\newcommand{\beq}{\begin{equation}}
\newcommand{\eeq}{\end{equation}}
\def\la{\hbox{\raise.35ex\rlap{$<$}\lower.6ex\hbox{$\sim$}\ }}
\def\ga{\hbox{\raise.35ex\rlap{$>$}\lower.6ex\hbox{$\sim$}\ }}
\def\runit{\hat {\bf  r}}
\def\phunit{\hat {\bfo \bvarphi}}
\def\zunit{\hat {\bf z}}
\def\beq{\begin{equation}}
\def\eeq{\end{equation}}
\def\beqa{\begin{eqnarray}}
\def\eeqa{\end{eqnarray}}
\def\sub#1{_{_{#1}}}
\def\order#1{{\cal O}\left({#1}\right)}
\newcommand{\sfrac}[2]{ \mbox{$\frac{#1}{#2}$}}
%
%
%
%



\title{Linear dynamics of weakly viscous accretion disks:
A disk analog of Tollmien-Schlichting waves.}

\author{O.M. Umurhan$^{1,2,3}$,
 and G. Shaviv$^{4,5}$\thanks{Email: gioras@physics.technion.ac.il}}

\offprints{O.M. Umurhan, \email{o.umurhan@qmul.ac.uk}}

\institute{
$^{1}$Astronomy Unit, School of Mathematical Sciences, Queen Mary
   University of London, London E1 4NS, U.K. \\
   $^{2}$Department of Geophysics and Space Sciences, Tel-Aviv University,
     Tel-Aviv, Israel\\
$^{3}$Astronomy Department,
City College of San Francisco, San Francisco, CA 94112, USA \\
$^{4}$Department of Physics, Technion-Israel Institute
of
Technology, 32000 Haifa, Israel\\
$^{5}$Institute of Theoretical Astrophysics,
 University of Heidelberg, 69120 Heidelberg, Germany}

\date{Received ---- / Accepted ----}

\titlerunning{Disk analog of Tollmien-Schlichting waves}

\abstract
{This paper discusses new perspectives and approaches to
the problem of disk dynamics
where, in this study, we focus on the
effects of viscous instabilities influenced by
boundary effects.
The Boussinesq approximation of the viscous large shearing box
equations is analyzed in which the azimuthal length scale of the
disturbance is much larger than the radial and vertical scales.
We examine the stability of
a non-axisymmetric potential vorticity mode, i.e. a PV-anomaly.
in a configuration in which buoyant convection
and the strato-rotational instability do not to operate.
We consider a series of boundary conditions which show
the PV-anomaly to be unstable both on a finite and semi-infinite radial domains.
We find these conditions leading to an instability which is the disk analog of
Tollmien-Schlichting waves. When the viscosity is weak, evidence of the instability
is most pronounced by the emergence of a vortex sheet at the critical layer
located away from the boundary where the instability is generated.
For some boundary conditions a necessary criterion
for the onset of instability for vertical wavelengths
that are a sizable fraction of the layer's thickness and
when the viscosity is small is that
the appropriate Froude number of the flow be greater than one.
This instability persists if more realistic boundary conditions
are applied, although the criterion on the Froude number
is more complicated.  The unstable waves studied here share qualitative
features to the instability seen in rotating Blasius boundary layers.
The implications of these results are discussed.  An overall new strategy
for exploring and interpreting disk instability mechanisms
is also suggested.

\keywords{accretion, accretion disks -- instabilities --}}
  \maketitle


\section{Introduction}\label{Introduction}

The magneto-rotational instability (Balbus, 2003, MRI hereinafter)
is generally considered to be the leading candidate explaining the
source of enhanced transport observed for disk systems.
Three conditions are required for its operation: the concurrent
presence of rotation and shear,
a primordial (no matter how small)
magnetic field, and sufficient ionization of the
fluid so that the gas is in the MHD regime.
Consequently, it is natural to ask the question:
What happens in an accretion disk if one or more of these conditions are not satisfied?
\par
Supposing that there are disks in which the MRI or any other
MHD (dynamo) mechanism is either weakly operating or entirely absent:
what else can drive activity, possibly even leading to turbulence?
Attempts to answer this question
include, but are not limited to, defects in the Keplerian profile (Li et al. 2000),
baroclinic instabilities (Klahr \& Bodenheimer, 2003, Johnson \& Gammie, 2005
Petersen et al. A-B, 2007),
transient growth and sustained subcritical dynamics (Richard \& Zahn, 1999, Iounnou \& Kakouris, 2000,
Chagelishvili et al. 2003, Tevzadze et al. 2003, Yecko
2004, Umurhan \& Regev, 2004, Umurhan et al. 2006, Barranco \& Marcus, 2005,
Lesur \& Longaretti, 2005, Lithwick, 2007,
and see the experiments of Richard, 2001 and Ji et al., 2006)
and, unsteady wave dynamics like
the Papalouizou-Pringle instability (Papalouizou \& Pringle, 1984),
hereinafter ``PPI",
and the strato-rotational instability (Dubrulle et al., 2005,
Shalybkov \& R\"udiger, 2005, Umurhan, 2006, Brandenburg \& R\"udiger, 2006), hereinafter
``SRI".  The majority of these recent investigations (excepting Yecko, 2004,
and Afshordi et al. 2005) have focused on strictly inviscid processes.
Indeed, the classical approach to such questions is to investigate
processes which might lead to turbulent transport by first establishing
a mechanism of linear instability from the vantage point of purely
inviscid (or nearly inviscid) flow.

Since astrophysical fluids have some effective viscosity - however small it may be -
we pose the question:
could a weakly viscous flow in a sheared and
rotating environment undergo an intrinsically viscous
type of linear instability that nonlinearly saturates with significant
amplitude?
A few previous studies have addressed this question.
Kato (1978) demonstrated that if a fluid's viscosity is a function
of the state of the fluid then a disk can experience pulsational
dynamics in a way similar in quality to stellar pulsation like
that in the theory of Cepheid variables.  Hereafter we shall
refer to this effect as the viscous pulsational instability (VPI).
Latter \& Ogilvie (2006) reexamined the VPI by studying
how even axisymmetric f-modes,
in a shearing sheet environment, create fluctuating
stresses that explicitly draw energy from the shear which
leads to overstability.
Kleiber \&
Glatzel (1999) have
shown that accretion tori (in this case, ones which have a
constant specific angular momentum distribution) can be linearly
unstable above a minimum Reynolds number.  Dubrulle et al. (2005)
and Shalybkov \& R\"udiger (2005) also report that the growth rate
of the SRI may be, under certain conditions,
enhanced by viscosity.\par
Before proceeding we present some remarks concerning the nature
of the fluid state to be studied as well as a discussion about
boundary conditions and a proposed alternative way to consider
their uses.
\subsection{Viscous considerations}

The non-commutative nature of the Navier-Stokes equations
in the limit of Re$ \rightarrow \infty $ with the Euler equations
is a long standing fact (see the discussion in Schlichting \& Gersten 2001, pg 96-8).
For instance, viscous stresses do not necessarily vanish on
the boundaries of a viscous flow when the viscosity limits to zero.
In turn this implies that one may not
properly take the Navier Stokes equations and na\"{i}vely substitute zero viscosity
to reach the inviscid limit (Lions 1993, Joseph 2003).
Aside from the generic appearance of boundary layers,
other effects can appear when a weak viscosity is
included into problems of study. For example,
in the nearly inviscid shallow-water theory
of strong shear flows (Balmforth, 1999),
normal-modes can emerge out of a continuous spectrum
when viscosity is introduced into the dynamics.
\par
Another example of the subtleties inherent in viscous flow is the instability
associated Tollmien-Schlichting waves (see discussion
of T-S waves in Schlichting \& Gersten, 2001, and Schmid \& Henningson, 2000) -
which are traveling waves appearing in wall-bounded flows
that neither grow nor decay in the
inviscid limit
and become unstable when viscosity
is included in the analysis (examples include plane
Pouiselle-flow and Blasius boundary layers).
Similar to the process discussed by Balmforth (1999), the
traveling mode becomes unstable through
the interaction of an inviscid normal mode and a viscous
normal mode - the latter of which exists only
as a member of the continuous spectrum in the inviscid
limit (Baines, Majumdar \& Mitsudera, 1996).
Far from being considered
mathematical oddities, T-S waves appear to play a prominent role
in the transition to turbulence in boundary-layer flows
(for a recent summary see Drazin, 2002).  A situation studied which
closely resembles the condition of an astrophysical disk is the
modal and non-modal response
of a rotating Blasius boundary layer (Yecko \& Rossi 2004) in which
instability is promoted
when the azimuthal scale of a disturbance is longer
than its vertical scale.

In a general sense, because the governing
equations are of a higher order in the viscous case,
a new  space of
possible solutions emerges which are either absent or inactive in the inviscid case.
Our main query is therefore: if rotationally supported flows are (linearly) well-behaved
in the {\em exactly} inviscid limit but the viscous flow shows some type of
dynamically significant behaviour  - even as the inviscid
limit is {\em approached} - then might it be misleading to test
stability only of exactly (Re $=\infty$) inviscid flows? Perhaps the
subtle nature of disks is linked to this feature.
That which best summarizes this perspective is the quote attributed
to the atmospheric dynamicist E. T. Eady where he is purported to
have said,
``It is not the process of linearization that limits insight.  It is the nature
of the state we choose to linearize about," (Bayley, Orszag \& Herbert, 1988).
\par
To prospect for an instability mechanism that might lead to
sustained unsteady behavior by assuming a turbulent viscosity model a
priori  might seem contradictory at first.  But, given
the difference in behavior known to exist (in other problems)
between nearly inviscid and exactly inviscid models it is
therefore mandatory to clarify these differences within the
context of an astrophysical disk too.
%
%
A conjecture that such an investigation could address is the following.
It is reasonable to suppose that disks are continuously fed
with a turbulent flow field either by in-fall or
some mass transfer processes.
Could this turbulent flow field undergo a secondary transition into another
dynamical state (possibly turbulence of a different stripe)
due to the turbulently enhanced viscosity?
Suggestions which hark on these lines of thought are
found in Doering et al. (2000), Kersal\'{e} et al. (2004)  and more directly in terms
of secondary transitions induced by Ekmann flow as suggested in
Lesur \& Longaretti (2005).  Overstability driven by material fluctuations
in the turbulent stresses (i.e. the VPI, Kato, 1978 and Latter \& Ogilvie, 2006)
are also candidates for such secondary transitions.

\subsection{Interpreting boundary conditions and their effects on dynamics}

Much attention has been devoted to evaluating processes which
are intrinisic in some way to the fluid - meaning to say
that it is assumed that it is more valuable to study those mechanical processes
which are minimally sensitive to
the boundary conditions imposed and maximally ``emerging" out of something essential
about the fluid and its basic state.  The MRI is
an example of this as well as other more basic fluid dynamical
instabilities such as the Rayleigh-Taylor and Rayleigh-Benard
instabilities.
We wonder if this approach to the question of disk turbulence may
be self-limiting
given that many fluid instabilities which lead to some form of
turbulence in terrestrial
flows are driven in large part by the boundary conditions of the system
(e.g. T-S waves and turbulent transition).
An alternative way is to view boundary conditions as a filter for
certain solutions or as a tool to classify solutions.
Kersal\'e {\em et} al. (2004) adopt a similar philosophy
by studying
the linear response of an incompressible fluid in a
Taylor-Couette type of cylindrical flow subject to a variety of boundary conditions.
Of course, the Taylor-Couette setup and the
boundary conditions they test are not what one would
expect in a terrestrial apparatus or experiment, however,
\emph{if taken as a metaphor for
a disk environment} then this sort of exploration allows one
to test, evaluate, understand and eventually categorize the dynamical
response of a fluid as a function of differing boundary
conditions.
\par
The inviscid PPI and SRI are examples of linear instabilities
which come about
due to the imposition of arguably artificial boundary conditions on
inner and outer walls of a model disk system.
On the other hand, these results may be intepreted in terms of the Hayashi-Young
criterion for wave instability which states that a physically separated
wave pair may promote linear instability if the waves
counterpropagate with respect to each other
with
nearly the same wavespeed and if the waves have an action-at-distance
effect upon each other
\footnote{By this it is meant to say that there exists
a wave evanescent region separating the waves in question.}
(Hayashi \& Young, 1987
and see also Sakai, 1989, Baines \& Mitsudera, 1994).
Indeed Goldreich, Goodman
and Narayan (1985)
\footnote{They show that the interior
of the domain
does not intrinsically support propagating waves as it behaves
like a wave evanescent zone. However the imposition
of the boundary conditions brings into existence waves that
propagate along the radial boundaries of the domain.}
 point out that the PPI may be viewed
as a process resulting from the interaction of a pair of
{\em edgewaves} mutually interacting with each other across
a wave-evanescent region
\footnote{
Examples of this process are well known in atmospheric flow
(Charney \& Stern, 1965, Hoskins et al., 1985, Davies \& Bishop, 1994).}.
The SRI may also be similarly rationalized (Umurhan, 2008).
\par
Thus although the counterpropagating edgewaves
responsible for the PPI and SRI are understood to result
from the use of unrealistic boundary conditions,
it is certainly
not ruled out that the general counterpropagating wave mechanism
could be at work
in real disks.
The linear instability of disks with two or more (potential)
vorticity defects (e.g. Li et al., 2000)
could be interpreted
as an instance of this process.

\subsection{An overview of the findings in this study}
From
a systematic asymptotic scaling analysis we derive in Section 2 and Appendix A
the equations appropriate to a box section of a viscous shearing
accretion disk (assuming an
$\alpha$ viscosity formalism)  by exploiting the smallness
of the parameter $\varepsilon$ which assesses the ratio
of the soundspeed to the rotation speed measured
at some radial point of a circumstellar disk.
We refer to this model as the Viscous Large Shearing Box (VLSB)
and these equations have appeared before (cf. Latter \& Ogilvie, 2006).
We are reminded
that while the velocity fluctuations in the shearing box
are an order $\varepsilon$ smaller than
prevailing rotational (``Keplerian") velocities,
the steady accretion velocities implied by
the alpha viscosity model are
an order $\varepsilon^2$
smaller than the same disk rotational velocity.\par
We consider the fate of a non-axisymmetric potential vorticity disturbance
(or simply ``PV-anomaly") subject to varying boundary conditions.
Accordingly, in Section 3 and 4 the VLSB are
analyzed in the limit where the perturbation's
azimuthal length scale is asymptotically larger than its radial
and vertical scales (i.e. the quasi-hydrostatic semigeostrophic
limit, Umurhan, 2006, QHSG for short).
Additionally we
assume that the vertical component of gravity and
entropy gradient are constants.
\par
In Section 4.1 we formulate energy integrals of this reduced system in order
to better understand what
can contribute to destabilizing the PV-anomaly .  The energy budget
is characterized by a Reynolds-Orr type of equation whose sources
and sinks are given by the energy which the PV-anomaly can
extract from the shear, receive from the boundaries or lose due to dissipation.
\par
In the spirit of
Kersal\'{e} et al. (2004) we analyze the response
of the PV-anomaly subject to a controlled array of boundary conditions.
In discussing boundary conditions
we refer to the boundary closest to central object as {\em starside}
as opposed to the side furthest away from the object to
which refer to as {\em farside}.
We consider the dynamics as occurring on both a semi-infinite
domain (farside at infinity) and on a finite domain. Below we summarize the
main findings. Note that we have made sure to consider boundary
conditions which
filter out the SRI or PPI instabilities.
 \par In Section 4.2 an asymptotic analysis is done for the limit where the scaled turbulent viscosity
parameter (defined in the text as $\epsilon$)
is small.
We find instability
if the Froude number of the flow exceeds 1 for modestly large
vertical wavenumber.
Additionally, the PV-anomoly interacts with a critical layer of the
flow
creating a potential vorticity sheet sheet
whose radial extent is $\epsilon^{1/3}$ the size of
the vertical extent of the disk.
This analysis illustrates how an inviscid edgewave phenomenon (due to
the no-normal flow starside boundary condition) becomes unstable
when viscosity is included.
Most importantly
is that the instability is driven in part by the injection of
energy through the boundary.
\par
 We consider in Section 4.3 finite domain
disturbances of the PV-anomaly and let the viscosity
parameter be an order 1 quantity.  The fourth order normal mode problem
requires us to appeal to numerical computational
methods for solutions.
We impose on the farside boundary that
both the disturbance pressures and PV-anomalies vanish.
At the starside boundary we require that there be no-normal
flow there (as above).  The remaining starside condition takes on
four possibilities: (a) the flow is rigidly coupled at the wall, (b) the perturbations
are stress-free, (c) the PV-anomaly is zero, (d) the PV-anomaly gradient is zero.
The first two of these conditions are physically realistic.  The latter two
offer a means to consider the the effect of energy injection (or lack thereof)
through the boundaries
and to compare with the analytical analysis.  For rigid and
stress-free boundary conditions we see clear indications of a Tollmien-Schlichting type
of instability,
similar to the instability of rotating Blasius boundary layers (Yecko \& Rossi, 2004) and the energy budget of the disturbances show that this process
does not draw upon energy across the boundaries.

\section{Viscous Large Shearing Box and its QHSG approximation}
In Appendix A we consider a box section of an $\alpha$-disk
centered about its midplane and at a distance $R_{_0}$ from the
central object.  If the disk is cold, then it means that
the quantity defined by the ratio of the typical value
of the local midplane disk soundspeed, $c_s$,
to the local Keplerian velocity, $V_0$,
\[
\varepsilon \equiv \frac{c_s}{V_0},
\]
is less than $1$ by some substantial amount: protoplanetary disks,
for example, are believed to have an $\varepsilon \approx 1/20$.
Using now familiar scaling arguments and exploiting the smallness
of $\varepsilon$ we derive
from the full equations of motion in cylindrical
coordinates appropriate equations of motion in what we refer to as the {\em
Viscous Large-Shearing Box} (VLSB for short).  The tactics and
procedures behind this effort are the same ones employed in the
derivation of the Large-Shearing Box (LSB) (Umurhan \& Regev,
2004) however, the viscous stresses are included.  We have,
\beqa  & & (\partial_t - q\Omega_0x
\partial_y)\rho + \nabla\cdot (\rho_b+\rho)\bf u'
= 0, \label{lsb_continuity}\\
& &  (\partial_t - q\Omega_0x \partial_y) u' + {\bf u'} \cdot\nabla
u' -2\Omega_0 v' = \nonumber \\
& &
\hskip 5.0cm
-\frac{\partial_x p}{\rho_b + \rho} + N_r'
\label{lsb_full_radial}\\
& & (\partial_t - q\Omega_0x \partial_y) v' + {\bf u'} \cdot\nabla
v' + (2-q)\Omega_0u' =  \nonumber \\
& & \hskip 5.0cm -\frac{\partial_y p}{\rho_b + \rho} +
N_\phi',
\label{lsb_azimuthal} \\
& & (\partial_t - q\Omega_0x \partial_y) w' + {\bf u'} \cdot\nabla
w'  = -\frac{\partial_z p + \rho g(z)}{\rho_b + \rho}  + N_z',
\label{lsb_vertical}\\
& &(\partial_t - q\Omega_0x \partial_y)\Sigma + {\bf u'}\cdot
\nabla \Sigma = 0 \label{lsb_entropy} \eeqa
in which the total
entropy is defined by
\[
\Sigma \equiv  \ln\frac{p_b + p}{\left(\rho_b +
\rho\right)^\gamma},
\]
and  $\gamma$ is the ratio of
 the specific heats at constant pressure to the specific heat at constant  volume.
The vertical component of gravity is dependent on $z$
\beq
g(z) = -\Omega_0^2 z. \label{vert_g}
\eeq
All primed quantities are perturbations about the basic flow.
 The viscous stresses are
 \beqa & & (\rho_b+\rho) N_r' = \tilde\eta(\partial_x^2 +
\partial_y^2)u' + \partial_z\tilde\eta\partial_z u' \nonumber \\
 & & \hskip 1.0cm  + \partial_x\left[\tilde\eta(\partial_x u' + \partial_y v') +
\partial_z\tilde\eta w'\right]
-\sfrac{2}{3}\partial_x\tilde\eta \nabla\cdot {\bf u}', \\
& & (\rho_b+\rho) N_\phi' = \tilde\eta(\partial_x^2 +
\partial_y^2)v' - q\Omega_0\partial_x\tilde\eta
+ \partial_z\tilde\eta\partial_z v' \nonumber \\
& & \hskip 1.0cm + \partial_y\left[\tilde\eta(\partial_x u' + \partial_y v') +
\partial_z\tilde\eta w'\right]
-\sfrac{2}{3}\partial_y\tilde\eta \nabla\cdot {\bf u}' , \label{N_phi}\\
& & (\rho_b+\rho) N_z' = \tilde\eta(\partial_x^2 + \partial_y^2)w' +
\partial_z\tilde\eta\partial_z w' + \nonumber \\
& & \hskip 1.0cm + \tilde\eta\partial_z(\partial_x
u' + \partial_y v')+  \partial_z\tilde\eta\partial_z w'
-\sfrac{2}{3}\partial_z\tilde\eta\nabla\cdot {\bf u}',
\eeqa where
\beq \tilde\eta = \sfrac{2\alpha}{3\Omega_0}(p_b + p). \eeq
The above equations are non-dimensional.  Time
is scaled by the local rotation time of the box.  All lengths
are scaled according to a length $H\ll R_{_0}$ {which is
comparable to the disk thickness (see Appendix A)}.  Pressures are
scaled according to the product of the
local midplane soundspeed and density, which is in
turn based on some fiducial characteristic temperature scale.  For
further details  see Umurhan \& Regev (2004). $x$ represents the
radial (shearwise) coordinate of the SB while $y$ is the azimuthal
(streamwise) and $z$ is the vertical coordinate (normal to the
disk midplane). The velocity components, i.e. ${\bf u'} =
\{u',v',w'\}$, are for the radial, azimuthal and vertical
directions.  It is important to keep in mind that these flow
variables represent {perturbations} about the steady Keplerian
flow. $\Omega_0$, sometimes also referred to as the Coriolis
parameter, {\em is $1$ in these nondimensionalized units}, meaning
to say because time has been scaled according to the dimensional
value of the rotation rate at $R_0$, i.e. $\Omega(R_0)$, the local
Coriolis parameter formally is equal to one. We retain this symbol
in order to flag the Coriolis effects in this calculation. The
local shear gradient is defined to be \beq q \equiv
-\left[\frac{R}{\Omega}\left(\frac{\partial \Omega}{\partial
R}\right)\right]_{R_0}, \eeq in which $\Omega(R)$ is the full disk
rotation rate. For Keplerian disks the value of $q$ is $3/2$. The
local Keplerian flow is represented here by a linear shear in the
azimuthal direction, i.e. $-q\Omega_0 x {\bf \hat y}$.\par The
steady state quantities are denoted with index b and in particular
we assume that pressure ($p_{_b}$) and density ($\rho_{_b}$)
profiles satisfy the hydrostatic balance relationship \beq
\partial_z p_{_b} =- \rho_{_b} g(z).
\label{basic_state_HE} \eeq
All corrections to this equations are
of a higher order and ignored here.
The expression responsible for the VPI can
be identified as fluctuating viscosity parameter in (\ref{N_phi}),
term $-q\Omega_0\partial_x\tilde\eta$.
\par
{We call to attention that the accretion and
meridional velocities characterizing $\alpha$-disks (Klu\'zniak \&
Kita, 1999) do not appear
in the VLSB equation set (see Appendix A).
By definition, turbulent disks exhibit accretion
velocities as they are the natural consequence of equations
describing global dynamics. The shear velocities are in fact quite
complex as Klu\'zniak \& Kita (1999) showed for the particular
case of an Shakura-Sunyaev type of $\alpha$-disk. The radial
velocities in steady state are found to be sheared in the vertical
direction and, as well, there exists a vertical component to the
flow with both radial and vertical dependence (the meridional
flow). However the scaling arguments implemented to reach these
``shearing box" equations, especially the relative scaling
relationships between the dynamical velocities and the accretion
scalings, show that the influence of the steady accretion rate
appears at higher orders in the expansion procedure.  In other
words, dynamical perturbations on the scale of the box do not feel
the effects of steady accretion and meridional flow - they only
feel the effects of the steady Keplerian shear.
 The scaling analysis also shows that
 the $\alpha$-viscosity (which is the driver of the accretion flow)
 does influence the dynamics at these scales and is the reason why it
 appears in these equations.}

\section{The quasi-hydrostatic semigeostrophic approximation of the VLSB equations}
The equations of motion may be simplified for further analysis by
implementing the quasi-hydrostatic semigeostrophic (QHSG) scaling
arguments used in Umurhan (2006). The QHSG is
useful in its ability
to expose the essential mathematical features of the
inviscid-SRI (Dubrulle et al., 2005, Umurhan, 2006).
\par
We suppose that the azimuthal scales of motion are much larger
than the radial or vertical scales.  We measure this with the
small parameter $\delta$.  In order to maintain
asymptotic validity we
assume the following orderings
\beq
\varepsilon \ll \delta \ll 1.
\eeq

Thus we suppose that the following operations upon dynamical
quantities scale accordingly as
\[
\partial_x, \partial_z \sim \order 1, \qquad \partial_y \sim\order\delta.
\]
Then we suppose that the radial and vertical velocities are
correspondingly smaller than the azimuthal velocities by this same
scale, in other words
\[
v' \sim \order 1, \qquad u',w' \sim \order \delta.
\]
These scalings say then that
\[
-qx\Omega_0\partial_y + u'\partial_x + v'\partial_y + w'\partial_z
\sim \order{\delta}.
\]
These scalings will make it easy to follow waves propagating with
respect to the background Keplerian flow velocity.  Therefore, the
temporal dependence should also scale by the scaling appropriate
to $\partial_y$.  It follows that
\[
\partial_t \sim \order \delta.
\]
Furthermore we say that the density, pressure (and by
implication, the entropy) fluctuations are all order 1, that is
\[
\rho,p,\Sigma \sim \order 1.
\]
The new issue that must be addressed here is to suggest a scaling
that brings in the viscous terms at the lowest non-trivial order.
To this end setting $\alpha\sim\order{\delta}$ achieves this goal
and we shall formally write $\alpha = \delta\alpha_1$.
In sum, then, to lowest order we have the following reduced set:
\beqa & & (\partial_t - q\Omega_0x \partial_y)\rho
+ \nabla\cdot(\rho_b+\rho)\bf u'
= 0, \label{lsb_continuity_qhsg}\\
& &
0 = 2\Omega_0 v'  -\frac{\partial_x p}{\rho_b + \rho},
\label{lsb_radial}\\
& & (\partial_t - q\Omega_0x \partial_y) v' + {\bf u'} \cdot\nabla
v' + (2-q)\Omega_0u' = \nonumber \\
& & \hskip 5.0cm -\frac{\partial_y p}{\rho_b + \rho} + N_y',
\label{lsb_azimuthal_qhsg}\\
& &
0 = -{\partial_z p} - {\rho g(z)},
\label{lsb_vertical_qhsg}\\
& & (\partial_t - q\Omega_0x \partial_y)\Sigma + {\bf u'}\cdot
\nabla(\Sigma_b + \Sigma) = 0, \label{lsb_entropy_qhsg} \eeqa
where we have introduced the basic state entropy $\Sigma_b$ and
its dynamically varying counterpart $\Sigma$ which are defined by
\beq \Sigma_b = \ln{\frac{p_b}{\rho_b^\gamma}}, \qquad \Sigma =
\ln{\frac{1+\frac{p}{p_b}}{\left(1+\frac{\rho}{\rho_b}\right)^\gamma}}.
\label{entropy_definition} \eeq
Only the azimuthal direction
stress component survives at lowest order due to this scaling argument,
\beq (\rho_b +
\rho)N_y' = \tilde\eta\partial_x^2 v' -q\Omega_0 \partial_x\tilde\eta
+ \partial_z\tilde\eta\partial_z v'.
\label{N_yprime}
\eeq
Although we have invoked scaling argments leading to the
above sets of equations we have not formally rewritten all
of the variables to signify these assumptions as it is our
desire to
preserve the transparency of the subsequent presentation.
Note that effect responsible for the VPI survives this scaling
argument as it appears in (\ref{N_yprime}) as the term $-q\Omega_0 \partial_x\tilde\eta$.

\section{Boussinesq Simplification, Assumptions and Linearized Dynamics}

In Umurhan (2006) it was shown that the QHSG approximation
of Boussinesq disk models recovers
the linearized hydrodynamic
behavior contained therein for concurrent small values of the azimuthal
wavenumber and wavespeed.
It was further demonstrated that the dynamics contained in the
QHSG approximation of the LSB is faithfully represented if one considers
instead the equivalent incompressible Boussinesq (Spiegel \& Veronis, 1960)
version of QHSG approximated LSB equations.
Applying this sequenced reasoning to the linearized
version of (\ref{lsb_continuity_qhsg}-\ref{lsb_entropy_qhsg}) gives,
\beqa
\partial_x  u + \partial_y v + \partial_z w &=& 0,
\label{Nsc_incompressibility} \\
0 &=& 2 w   - \partial_x \Pi,  \label{Nsc_u_lin}
\\
\left(\partial_t -  qx \partial_x\right) v +
(2-q) u &=& -\partial_y \Pi + \tilde N_y', \label{Nsc_w_lin}
\\
0 &=&
-\partial_z \Pi + \Theta,
\label{Nsc_v_lin} \\
\left(\partial_t -  q x \partial_x\right) \Theta
&=&  - N^2 w
, \label{Nsc_theta_lin}
\eeqa
where we have dropped all primes from the velocity quantities
and explicitly set $\Omega_0$ to its value of 1.
In the usual Boussinesq approximation,
density fluctuations are dynamically significant
when coupled to gravity.  In these circumstances $\rho$
is replaced by $-\theta$.
The non-dimensionalized temperature quantity $\theta$ and
its associated steady state temperature field $T_b(z)$ are characterized
by the (linearized) conservation relation,
\[
(\partial_t - qx \partial_y)\theta +w\partial_zT_b = 0.
\]
For clarity we  reexpress
this thermal quantity in terms of $\Theta$ given by $ \Theta
\equiv g  \theta/ \rho_b$
\footnote{A momentary comparison to (\ref{lsb_entropy_qhsg})
should convince the reader that
$\Theta$ represents a perturbed entropy
quantity (Dubrulle et al., 2005) making
(\ref{Nsc_theta_lin}) a reasonable analog of the
linearized form of (\ref{lsb_entropy}).}.  In the
Boussinesq approximation fluctuating density variables
influence the dynamics when coupled to gravity.
Forthwith, $\rho_b$ and $P_b$ are
taken to be constant (set to 1)
and, furthermore, $\rho_b$ is
absorbed into the fluctuating pressure leading to defining the enthalpy $\Pi \equiv
p/\rho_b$.  All unprimed velocities (i.e.
$u,v,w$) are now understood to represent linearized disturbances.
The non-dimensionalized {\em Brunt-V${\ddot { a }}$is${\ddot { a }}$l${\ddot { a }}$
frequency}, $N$, emerges in the equation for the {perturbation}
temperature field (\ref{Nsc_theta_lin}) and is given by \beq N^2
\equiv g \sfrac{1}{\rho_b}
\partial_z T_b. \label{def_BV_freq} \eeq Throughout this study $N$
is taken to be real (buoyantly stable). The azimuthal stress is
\beq \tilde N_y'
=\sfrac{2\alpha_1}{3}\left[\left(\partial_x^2
+\partial_z^2\right)v -
q\partial_x\Pi\right].
\eeq
The term that gives rise to the VPI
appears in the above as $\sim-q\partial_x\Pi$.
We proceed further by (i) operating on (\ref{Nsc_w_lin}) with
$\partial_x$, (ii) operating on (\ref{Nsc_u_lin}) with
$\partial_y$, (iii) and subtracting the results to reveal
\beq
(\partial_t - q x\partial_y)\partial_x v = (2-q)\partial_z w + \partial_x N_y',
\eeq
where the incompressibility condition was used in writing the first term on the RHS of
this expression.
Multiplying (\ref{Nsc_theta_lin}) by $(2-q)/N^2$
followed by operating on the result with $\partial_z$
gives
\beq
(\partial_t - q x\partial_y) (2-q) \partial_z (\Theta/N^2)
= -(2-q)\partial_z w.
\eeq
Adding these two equations together and making use of the relationships
(\ref{Nsc_u_lin}) and (\ref{Nsc_v_lin}) yields the following single equation for $\Pi$:
\begin{eqnarray}
& &  \left({\partial_t} - q x{\partial_y}\right)
\left({\partial_x^2} + \partial_z F_\epsilon^2
\partial_z\right)\Pi
= \nonumber \\
& & \hskip 3.0cm \frac{1}{{\rm Re}}\left(\partial_x^2 + \partial_z^2\right)
\partial_x^2\Pi - \frac{1}{{\rm Re}}2q\partial_x^2\Pi.
 \label{the_equation_of_motion}
\end{eqnarray}
To showcase the elements in the above we have written
$2\alpha_1/3 \leftrightarrow {\rm Re}^{-1}$ in order
to remind ourselves that the $\alpha_1$ parameter is like the inverse of a Reynolds Number.
\footnote{However, note that the actual Reynolds number, denoted
by R, is a factor of $1/\delta$ larger than Re, i.e. R = Re$/\delta$.}
We have also defined the epicyclic Froude number
\[
F_{\epsilon}^2 \equiv \frac{2(2-q)}{N^2},
\]
which is in general a function of the vertical coordinate $z$
and vanishes on the symmetry axis driving the local Froude number
to very large values.
As we shall see, $F_{\epsilon}$ determines the character of the solutions
that emerge.
Umurhan (2006) demonstrates by a direct comparison of calculations that the onset of the SRI
(in the QHSG limit of the LSB equations explored there)
is reasonably well captured analytically when $N^2$ and $g$ are assumed to be
constants.
Guided by these previous results as well as similar
use in a series of other studies (e.g. Tevzadze, et al., 2004, Bodo et al. A-B, 2007)
we shall assume
\beq
g = {\rm constant},\qquad
N^2 = {\rm constant},
\label{g_assumption}
\eeq
implying that $F_\epsilon$ is a constant as well, and we restrict our considerations
to vertically periodic solutions.  We shall return to this matter in the Discussion.
\par
In this asymptotic theory
\[
Q \equiv (\partial_x^2 + \partial_z F_\epsilon^2\partial_z)\Pi
\]
corresponds to a the perturbed potential vorticity
(Tevzadze et al. 2004, Umurhan, 2006), also known
as the potential vorticity anomaly (Hoskins et al., 1985) and also called vortensity
in the astrophysical literature (e.g.
Klahr \& Bodenheimer, 2003).  Throughout the
rest of this work we will interchangeably use the terms
\emph{potential vorticity perturbation}, \emph{potential vorticity disturbance}
and \emph{potential vorticity anomaly} (i.e. PV-anomaly).
The PV-anomaly Q relates to a vorticity pointing in the
vertical direction.
\par
We consider travelling wave normal mode solutions to (\ref{the_equation_of_motion}):
the modes are assumed
to be
azimuthally periodic on scale $L_y$ and vertically periodic
on scale $L_z$,
\beq
\Pi = \hat\Pi(x)e^{ik(qct + y) + i\beta z} + {\rm c.c.},
\eeq
where the azimuthal wavevector $k$ can be any positive number while
the vertical wavevector $\beta$ is any real number.
The wavespeed $c$ can be complex:  when ${\rm Im}(c) < 0$ there is growth
of the wave.
Furthermore we define
\beq
\epsilon \equiv \frac{2\alpha_1}{3kq} = \frac{1}{{\rm Re}kq}, \label{epsilon_definition}
\eeq
to be a wavenumber scaling of the viscous parameter.  Thus the equation governing
the structure function $\hat\Pi(x)$ becomes
\beq
i(c-x)(\partial_x^2 - k_{_F}^2)\hat\Pi = \epsilon(\partial_x^2 - \beta^2- 2q)
\partial_x^2\hat\Pi.
\label{linear_eqn_motion}
\eeq
The Froude-wavenumber is defined as $k_{_F} \equiv F_\epsilon \beta$.
(\ref{linear_eqn_motion}) will be the fundamental equation of study.  Furthermore
expressed in this form, the azimuthal and radial velocities are
\beqa
 \hat v &=& \sfrac{1}{2}\partial_x\hat\Pi, \label{hat_w_def} \\
 \frac{2-q}{qk} \hat u &=& -\sfrac{1}{2}(c-x)i\partial_x\hat\Pi -i\left(\sfrac{1}{q}\right){\hat\Pi} \nonumber \\
& & \ \ \ \ \ \ \ \ +\sfrac{1}{2}\epsilon(\partial_x^2 - \beta^2 - 2q)\partial_x\hat\Pi,
\label{u_Pi_relationship}
\eeqa
with the vertical velocity following from evaluating the incompressibility equation
at the boundary and using (\ref{linear_eqn_motion}), i.e.
\beq
\hat w = \frac{qk k_F^2}{2\beta(2-q)}\Pi.
\eeq
Finally, the perturbed potential vorticity is $\hat Q \equiv (\partial_x^2 - k_F^2)\hat\Pi$.
The remainder of the boundary conditions, namely in the $x$ direction, will be stated
in the following sections according to the problem being solved.

\subsection{Energy Integrals}\label{global_integrals}
It is instructive to develop global energy integrals
as such quantities aid in developing an interpretation of
the results in the following sections.  We restrict our attention
to the energetics associated with the QHSG-Boussinesq model set
(\ref{Nsc_u_lin}-\ref{Nsc_incompressibility}) keeping in mind
the assumptions we made about vertical and azimuthal periodicity
and the constancy of $N^2$.
The radial conditions are left arbitrary
and they will be dealt with accordingly in each of the subsequent subsections.
We proceed by defining the perturbation
thermomechanical energy density ${\cal E}$
as
\[
{\cal E} = \frac{1}{2}\left(v^2 +\frac{\Theta^2}{N^2}\right),
\]
by multiplying (\ref{Nsc_w_lin}) by $w$ and (\ref{Nsc_theta_lin}) by
$\Theta/N^2$ and integrating over a domain which is periodic
in the vertical and horizontal directions, and finite in the
radial direction given by $0<x<x_1$, where $x_1$ is left
arbitrary, we find
\beq
\frac{dE}{dt} = \dot E_{{\rm shear}} + \dot E_{{\rm stress}}
-\dot E_{{\rm visc}},
\label{Reynolds_Orr}
\eeq
in which volume integrals are
\beqa
& & E \equiv \int_{\bf V} {\cal E} d^3{\bf x},\qquad
\dot E_{{\rm shear}} \equiv
 q\int_{\bf V} u v d^3{\bf x}, \nonumber \\
& & \ \  \dot E_{{\rm visc}}\equiv
\sfrac{2\alpha}{3}\int_{\bf V}
\left(|\partial_x v|^2 + |\partial_z v|^2 + 2q|v|^2
\right)d^3{\bf x},
\label{Volume_Integrals}
\eeqa
integrated on volume ${\bf V}$
with the volume element $d^3{\bf x} = dxdydz$ and where
the surface term is
\beq
\dot E_{{\rm stress}} \equiv
\int\left[-\Pi u + \frac{\eta}{\rho_b}v\partial_x v\right]_0^{x_{{\rm out}}} dydz.
\label{Surface_Integral}
\eeq
In writing $\dot E_{{\rm stress}}$ we have made use of the
periodicity conditions in the vertical and horizontal directions.  The bracketed
term as appearing means
\[
[f]^{x_{\rm out}}_0 \leftrightarrow f({x_{\rm out}}) - f(0),
\]
where $x_{{\rm out}}$ is the location of the outer boundary.
\par
The energy integral (\ref{Reynolds_Orr}) is the Reynolds-Orr Equation
appropriate for this QHSG system.
The energy $E$ is composed of the
baroclinic thermal term ($\sim \Theta^2$) plus a kinetic energy
term ($\sim w^2$), however, the kinetic energy term contains only the
azimuthal velocity contribution because the vertical and
horizontal velocity contributions are small by comparison
\emph{in the scaled system of equations} according to the QHSG
approximation of the original set (see Section 3 and Umurhan, 2006,
for details).  The term $-\dot E_{{\rm visc}}$ represents the
integrated losses and is comprised of
viscous losses due to the azimuthal velocity since, by the same reasoning
as above, the corresponding losses due to vertical and radial velocities
(in the scaled system) are
negligible.  The perturbed viscous stress term,
responsible for the VPI plays, a destabilizing role for
these PV-anomalies
as it appears in the above as the
the term
proportional to $2q|v|^2$.
The total external stresses on the system
is given by $\dot E_{{\rm stress}}$ and is comprised of
the surface integrated body pressure and the surface
viscous stress - the latter of which is expressed only
with the tangential stress due to $w$.  Finally
$\dot E_{{\rm shear}}$ is the Reynolds stress due to
the background shear state.  This expression may
be interpreted as accounting for the amount of
energy perturbations extract from the background
shear state.
We note also
that inspection shows that
$\dot E_{{\rm visc}}>0$ always, while
the remaining terms $\dot E_{{\rm stress}}$ and
$\dot E_{{\rm shear}}$ may be either positive
or negative given the state of the perturbed
flow or the boundary conditions employed.

\subsection{Asymptotic theory on a semi-infinite domain}\label{normal_modes_semi_infinite}

The following assumptions are made in order to proceed analytically:
(i) the domain in the $x$ direction lies between $0$ and $\infty$, thus
we have $x_{{\rm out}} = \infty$,
(ii) we require that all quantities decay as
$x\rightarrow \infty$,
(iii) there is no normal-flow at $x=0$ - this inner location is considered
{\em starside} as
it is closest to the central object,
(iv) the viscosity is weak but finite, hence, we assume that $\epsilon \ll 1$
\footnote{With respect to the other scalings we have
called on in this work, we shall formally require that
$\varepsilon \ll \delta \ll \epsilon \ll 1$.}
(v) even though there is viscosity in the problem we impose no particular
stress condition at $x=0$ and let the fluid quantities
be dictated by what emerges in the interior
of the domain.  This essentially means that flow stresses at
the starside are allowed to adjust according to the
dynamical response happening in the interior of the domain.
In practice it translates to only enforcing the no-normal flow
boundary condition.\footnote{Note, however, that we show in Section 4.3
that the results
obtained in this section, including the calculation
of the growth rates, would be essentially unchanged
had we imposed, instead, that the radial gradient of the
potential vorticity be zero at the starside boundary.}

\subsubsection{Expansions and outer solution}
We assume the following expansions well aware that this
is a singular perturbation calculation because of the presence of the critical
layer (see below).  The solution for the wavespeed $c$ is assumed of the form
\beq
c = c_0 + \epsilon c_1 + \cdots,
\eeq
and a similar series of the form
\beq
\hat\Pi = \hat\Pi_0 + \epsilon \hat\Pi_1  + \cdots
\eeq
We know aposteriori that the critical layer will generate a solution proportional to
$\epsilon\ln\epsilon$ and this is why such a term appears in the above expression.
Thus at $\order{1}$ we find that
\beq
(c-x)(\partial_x^2 - k_{_F}^2)\hat\Pi_0 = 0.
\eeq
To this order the no-flow boundary condition at $x=0$ amounts to
\beq
0 = c_0\partial_x\hat\Pi_0 + \frac{1}{q}\hat\Pi_0, \qquad {\rm at} \ \ x=0.
\label{order_0_bc}
\eeq
The solution to this equation which decays as $x\rightarrow\infty$ is
\beq
\hat\Pi_0 = A_0 e^{-k_{_F}x}, \label{hatPi0}
\eeq
where $A_0$ is an arbitrary amplitude.
Using this solution in boundary condition (\ref{order_0_bc}) amounts to selecting
$c_0$, which is
\beq
c_0 = \frac{2}{k_{_F}q} = \frac{2}{q F_\epsilon \beta}.
\label{PV_wavespeed}
\eeq
This says that the wavespeed is real and positive, which means in this case that
there will be a critical layer in the domain, i.e. at $x = x_c \equiv c_0$
(see below).\par
At $\order{\epsilon}$ we find the equation
\beqa
& & (c_0-x)(\partial_x^2 - k_{_F}^2)\hat\Pi_1 = \nonumber \\
& & \hskip 1.0cm -c_1(\partial_x^2 - k_{_F}^2)\hat\Pi_0
 -i(\partial_x^2 - \beta^2 - 2q)
\partial_x^2\hat\Pi_0.
\eeqa
Using (\ref{hatPi0}) for $\hat\Pi_0$ and dividing the equation by $c_0-x$ we find
the more transparent form
\beq
(\partial_x^2 - k_{_F}^2)\hat\Pi_1 = \frac{\Lambda A_0 e^{-k_{_F}x}}{i(x-c_0)}
+ C_1\delta(x-c_0).
\eeq
In which we have defined the parameter
$\Lambda \equiv k_{_f}^2(k_{_F}^2 - \beta^2 - 2q)$.  The delta
function appearing with the arbitrary coefficient $C_1$ is a formal device
used to signal the presence of a critical layer in the flow
(Case, 1960, and more recently,
Balmforth \& Piccolo, 2001, Balmforth et al., 2001).
The coefficient
will result from the critical layer analysis (see below) which follows
the procedures found in similar investigations (e.g. Stewartson, 1981,
Balmforth \& Piccolo, 2001).
In practice it means that we must separately develop solutions to $\hat\Pi_1$ on
either side of $x=x_c$.  Before doing so let us observe the reason why this
series expansion fails near $x_c$ by developing the solution to $\hat\Pi_1$ in
the vicinity of this point.  An indicial analysis shows that
\beqa
& & \hat\Pi_{1} \sim i^{-1}\Lambda A_0 e^{-2/q}\bigl[(x-c_0)\ln(x-c_0) - (x-c_0)\bigr] + \cdots \nonumber \\
& & \partial_x\hat\Pi_{1}  \sim i^{-1}\Lambda A_0e^{-2/q}\ln(x-c_0) + \cdots \nonumber \\
& & \partial_x^2\hat\Pi_{1}  \sim i^{-1}\Lambda A_0e^{-2/q}\frac{1}{x-c_0} + \cdots
\eeqa
These expressions show, especially that for
$\partial_x^2\hat\Pi_{1} $, that the solution begins to breakdown (i.e. break order)
when the quantity $|x-c_0|$ starts to approach $0$.  This divergence
must be controlled by considering a boundary layer calculation in and around the
critical layer.  Note that since the vertical vorticity is proportional
to $\partial_x^2\hat\Pi_{1}$, the critical layer will appear as
a vortex sheet.  The formal presentation of this solution, including the region of
validity and expression of the boundary conditions at this order is presented
in Section \ref{outer-solution-details}.


\subsubsection{Critical layer calculation, matching, and growth rate and analysis}
As we demonstrated above, the solutions begin to breakdown in the
vicinity of the critical layer which are those places where $x-x_c$ begins to
get small.  We must therefore reexamine (\ref{linear_eqn_motion}) in this
zone and to this end we define a new inner coordinate as
\beq
\epsilon^{1/3}\xi \equiv x-x_c.
\eeq
According to this new coordinate (\ref{linear_eqn_motion}) is reexpressed
as
\beqa
 \left(\frac{1}{i}\partial_\xi^4 - \xi\partial_\xi^2\right) \hat \Pi &=&
   \epsilon^{2/3}\left[
\frac{1}{i}(\beta^2 + 2q)\partial_\xi^2
+ \xi k_{F}^2\right]\hat\Pi \nonumber \\
& &  \hskip 1.0cm + \epsilon^{2/3}c_1\left[\partial_\xi^2
- \epsilon^{2/3} k_{F}^2 \right]\hat\Pi.
\label{critical_layer_eqn}
\eeqa
We intorduce a series expansion for the solution to
(\ref{critical_layer_eqn}) by writing
\beqa
\hat\Pi &=& \tilde\Pi_0 + \epsilon^{1/3}\tilde\Pi_{1/3}
+ \epsilon^{2/3}\tilde\Pi_{2/3} \nonumber \\
& & \hskip 1.5cm + \epsilon\tilde\Pi_{1}
+ \epsilon \ln \epsilon \tilde{\cal P}_1
+ \epsilon^{4/3}\tilde\Pi_{4/3}
+\cdots
\label{critical_layer_solution_expansion}
\eeqa
The remainder of this calculation including the matching of the
inner and outer solutions and the determination of the growth rate
$c_1$ has been relegated to Appendix \ref{critical-layer_calculation}. {We note that the term
proportional to $ \epsilon \ln \epsilon$ is needed for matching
purposes as the inner solution is extended out of the critical
layer (for details see the full exposition in the Appendix).} We
find that the correction wavespeed obeys
\beqa c_1 &=&
-\frac{2\pi}{q}e^{4/q}\Bigl(k_{_F}^2 -\beta^2-2q\Bigr)
\nonumber \\
& &-i\biggl(k_{_F}^2
 -
\beta^2 -2q\biggr)\left[1+\frac{2}{q}e^{-4/q}{\rm
Ei}\left(\frac{4}{q}\right) \right] .
\label{c1_correction} \eeqa
The growth
rate of this mode, i.e. $\sigma =  -\epsilon qk $Im$(c_1)$, is proportional to
\beq
\sigma =
{\rm Re}^{-1}\biggl((F_\epsilon^2-1)\beta^2 - 2q\biggr)
\left[1+\frac{2}{q}e^{-4/q}{\rm Ei}\left(\frac{4}{q}\right)\right],
\label{c1_correction_krho0}
\eeq
where we have restored the definition of $\epsilon$ in terms of Re,
and where we have replaced $k_{_F}$ accordingly with $F_\epsilon \beta$.
The term
inside the square brackets is always greater than zero for $q>0$
and limits to $3/2$ as $q\rightarrow 0$.\footnote{Note, however, this limit breaksdown
the QHSG approximation and is not considered.}
Therefore there is
growth when $F_{_\epsilon}^2 > 1 + 2q/\beta^2$.\par
The instability emerges from the
inner boundary due to application of the no-normal flow condition
but it will be at the critical layer where evidence of it
appears in the form of a pronounced PV-anomaly.
The radial width of this vortex zone is proportional
to $\epsilon^{1/3}$ the size of the box and the amplitude will
be $\epsilon^{2/3}$ times the leading order perturbation pressure field
(see the end of  Appendix \ref{critical-layer_calculation}).  Restoring units and
recalling that the shearing-box has been scaled according to the
thermal scale height of the disk $H$ , the radial
width of this vortex zone, $\Delta R$, is
\[
\Delta R\sim \alpha^{1/3} H.
\]
\par
Further analysis of these solutions together
with the Reynolds-Orr equation (\ref{Reynolds_Orr}) and
the definitions (\ref{Volume_Integrals}-\ref{Surface_Integral}) shows
that to lowest order in Re$^{-1}$
\beqa
\dot E_{{\rm stress}} &=& {{\rm Re}^{-1}} 2 k_F\beta^2 F_\epsilon^2 + \order{{\rm Re}^{-2}},\nonumber \\
\dot E_{{\rm visc}} &=& {{\rm Re}^{-1}}k_F\beta^2(F_\epsilon^2 +1 + 2q/\beta^2)  + \order{{\rm Re}^{-2}}
\eeqa
so that
\[
\dot E_{{\rm stress}} - \dot E_{{\rm visc}} =
{{\rm Re}^{-1}}k_F\beta^2(F_\epsilon^2 -1 - 2q/\beta^2)  + \order{{\rm Re}^{-2}}.
\]
A general evaluation of the Reynolds stress term $\dot E_{{\rm shear}}$ shows that it
contributes first at order ${{\rm Re}^{-1}}$ as well, i.e.
\beqa
\dot E_{{\rm shear}} &=& {{\rm Re}^{-1}}\dot E_{{\rm shear}}^{(1)}
+ \order{{\rm Re}^{-2}},  \nonumber \\
\dot E_{{\rm shear}}^{(1)} &=&
\int_{\bf V}(\hat u_1 \hat v_0^* + \hat u_0 \hat v_1^* + {\rm c.c.})d^3{\bf x}.
\nonumber
\eeqa
$\dot E_{{\rm shear}}^{(1)}$ is the first order correction to the
azimuthal/radial velocity correlations.  Its details may be worked out
but we leave it here in general form in order to make the following argument.
The Reynolds-Orr equation evaluated for this
set of boundary conditions is to lowest order in Re$^{-1}$,
\beqa
\frac{dE}{dt} \sim 2\sigma{\hat{\cal E}} &=& {\rm Re}^{-1}\left[ k_F\dot E_{{\rm shear}}^{(1)}
+\beta^4(F_\epsilon^2 -1- 2q/\beta^2)\right]
,\nonumber \\
{\hat{\cal E}} &=& \sfrac{1}{2}\beta^2\left[\sfrac{1}{4}F_\epsilon^2 + 1 \right] > 0,
\label{Reynolds_Orr_semiinfinite}
\eeqa
plus a correction which is $\order{{\rm Re}^{-2}}$.
In the case of instability, i.e. $F_\epsilon^2 > 1 + 2q/\beta^2$,
it follows that $\dot E_{{\rm stress}} - \dot E_{{\rm visc}}<0$.
Given
(\ref{c1_correction_krho0}) taken together with
(\ref{Reynolds_Orr_semiinfinite}) we find
that under conditions
of linear instability it also follows that
\beqa
k_F\dot E_{{\rm shear}}^{(1)} &=& \beta^4 (F_\epsilon^2 -1 - 2q/\beta^2)
\times \nonumber \\
& & \left[\left(1+\frac{1}{4}F_\epsilon^2\right)
\left(1 + \frac{2}{q}e^{-4/q}{\rm Ei}\left(\frac{4}{q}\right)\right)-1\right].
\label{explicit_E_shear}
\eeqa
Let us reflect upon this for a moment: with these boundary
conditions it is always the case that $\dot E_{{\rm stress}}-\dot E_{{\rm viscous}} > 0$
when $F_\epsilon^2 > 1 + 2q/\beta^2$.
The combined action of the starside viscous stresses and the domain integrated
viscous lossess still results in a net transfer of
energy into the domain.  Because
the term within the square brackets in (\ref{explicit_E_shear})
is always positive, these conditions
promote the type of corrections (at order Re$^{-1}$)
to the azimuthal and radial velocity profiles
such that a positive global correlation between them emerges and, hence,
resulting in $\dot E_{{\rm shear}}^{(1)} > 0$ when $F_\epsilon^2 > 1+2q/\beta^2$.
We conclude that with these boundary conditions the instability observed is
fed both by the energy entering the domain from the starside boundary
as well as by the energy extracted by the shear due to the resulting
order Re$^{-1}$
velocity profiles.

\begin{figure}
\begin{center}
\leavevmode \epsfysize=9.cm
\epsfbox{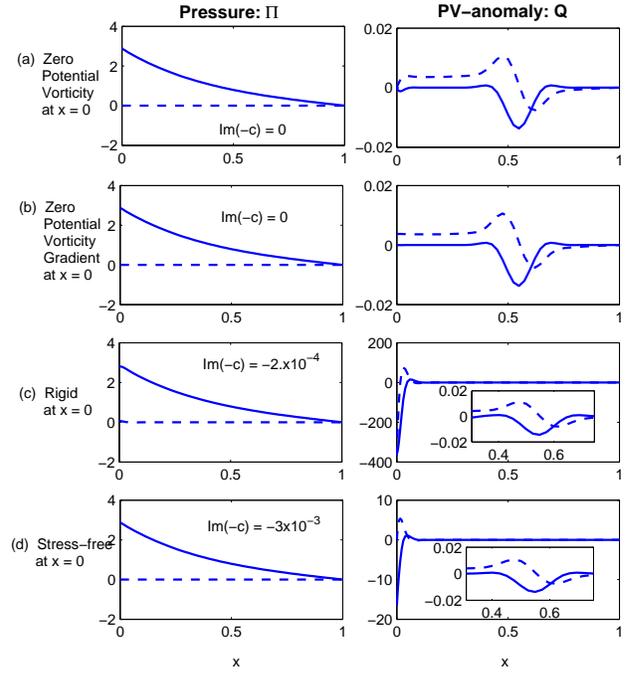}
\end{center}
\caption{{A comparison of eigenfunctions for a variety of starside
boundary conditions.  In all plots $\beta = 2$, $F_\epsilon = 1.2$ and $\epsilon = 10^{-4}$.
The pressure eigenfunctions $\hat \Pi$ are shown on the left
panel of plots while the potential vorticity eigenfunctions are shown
on the right panel.  The predicted growth rates are also quoted. (a)
Zero potential vorticity, (b) zero potential vorticity gradient, (c)
rigid boundary, (d) stress-free and (e) zero-pressure fluctuation.
.}} \label{bc_comparison_plot}
\end{figure}

\begin{figure}
\begin{center}
\leavevmode \epsfysize=9.cm
\epsfbox{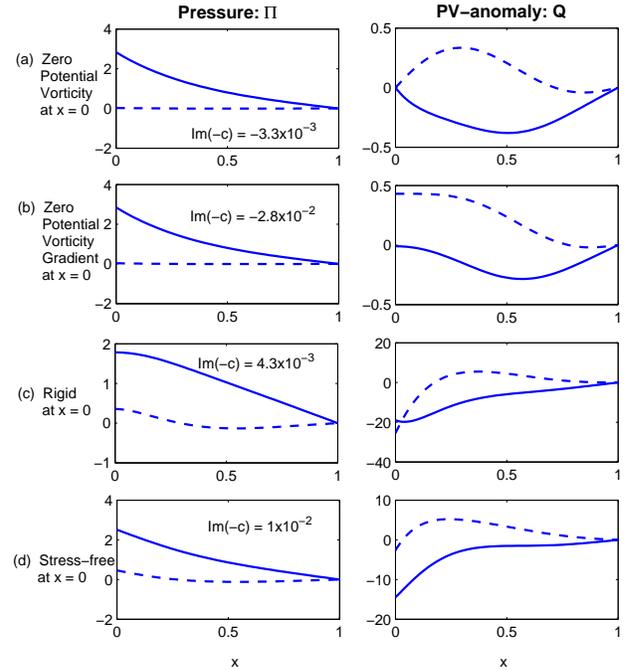}
\end{center}
\caption{{Like Fig. \ref{bc_comparison_plot} except $\epsilon = 10^{-2}$.
}} \label{bc_comparison_2nd_plot}
\end{figure}

\subsection{Finite radial domain investigations}
In this section we consider the normal-mode solutions
of (\ref{linear_eqn_motion}) occurring on a finite radial domain
where $x_{{\rm out}}=1$. All solutions
are computed numerically using a Newton-Raphson scheme on a Chebyshev
grid of anywhere from 33 to 129 points - higher
resolution is needed for smaller values of $\epsilon$.
All numerically generated
solutions are normalized so that $\int_{0}^1 \hat \Pi dx = 1$.
Because
this is a fourth order system we must specify four boundary conditions.
In all
of the following calculations two of the boundary conditions
will be that
\beq
\hat\Pi = 0, \qquad \hat Q = 0, \qquad {\rm at} \quad x=1,
\eeq
in other words, that the
pressure fluctuation and PV-anomaly are
zero on the farside boundary.  According to the normal-mode PV-anomaly,
$\hat Q \equiv \partial_x^2 \hat \Pi - k_F^2 \hat \Pi$, the fixed pressure condition
implies that at the far boundary $\partial_x^2\hat\Pi = 0$.  It therefore follows
from (\ref{hat_w_def}) that the perturbed stress expression, $\partial_x\hat v $,
is zero there as well.  A third condition will be that
there is no normal-flow on the starside boundary (as before),
i.e.
\beq
\hat u = 0, \qquad {\rm at} \quad x = 0.
\eeq
For the remaining starside boundary we shall explore four different conditions
enumerated in the corresponding subsections below.   {The most physically
plausible viscous starside condition is to set to zero either the azimuthal velocity fluctuation
or the azimuthal stress fluctuation.   We have also considered zero PV-anomaly and zero
PV-anomaly gradient conditions.  Although these conditions are less
physically realistic, they are simpler to interpret in terms of the energy arguments developed
in previous sections.}
\subsubsection{Rigid and stress-free starside boundary}
\begin{figure}
\begin{center}
\leavevmode \epsfysize=13.5cm
\epsfbox{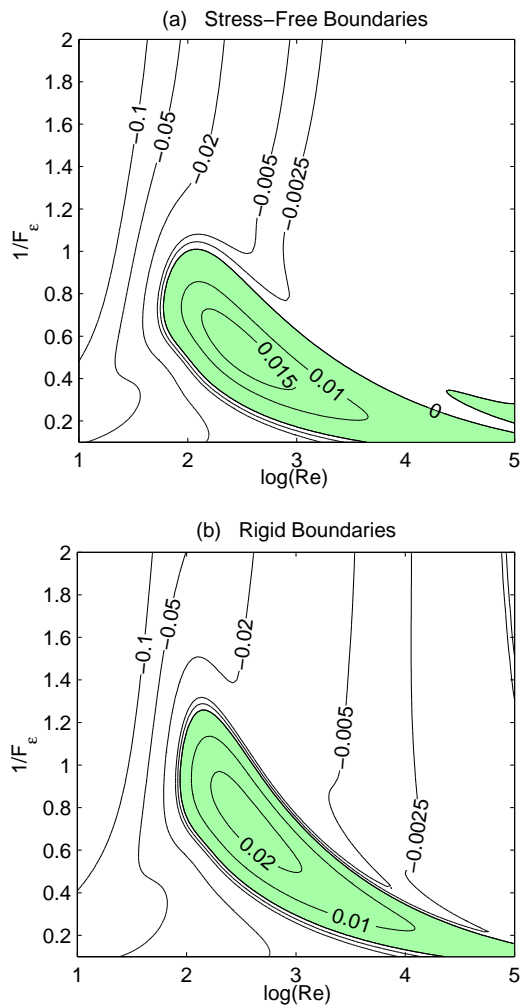}
\end{center}
\caption{{Contours on the $F_\epsilon$-Re plane of Im($-c$) for $q=3/2, k=1$ and $\beta = 2$.
Given the value of $k$, according to its definition $\epsilon$ is written in terms of Re $\equiv 3/(2\alpha)$:
(a) Stressfree boundary conditions at $x=0$, (b)
Rigid boundary conditions at $x=0$.  Shaded regions indicate growing modes.
Note that according to (\ref{PV_wavespeed}) the vertical axis indicates
increasing wavespeed.}}
\label{growthrates_rigidstressfree}
\end{figure}

These conditions translate to requiring $\hat v = 0$ (rigid) or $\partial_x \hat v = 0$
(stress-free) at $x=0$.    {Note that the usage of the term ``stress-free" to
describe the boundary condition
really refers to the perturbed part of the azimuthal flow as being stress-free.}
Inspection of Fig. \ref{growthrates_rigidstressfree}
shows that there are regions in
the $\epsilon$-$F_\epsilon$ parameter plane in
which there is instability
when rigid or stress-free conditions are imposed on
the starside boundary.
Although the parameter
range for instability is not as expansive as it is
for the other   {less-realistic} boundary conditions explored (see both
the previous section and below),
  {the results suggest that a rough
criterion for linear instability is $F_\epsilon > 1$.}
Furthermore, under these
starside conditions
we find that
the Reynolds-Orr equation (\ref{Reynolds_Orr})
simplifies to
\beq
\frac{dE}{dt} = \dot E_{{\rm shear}} - \dot E_{{\rm visc}},
\label{RO_no_vorty}
\eeq
since the zero-stress and rigid conditions
implies that $\dot E_{{\rm stress}} = 0$.  Thus we find that
the boundary conditions are such that instability is not
directly driven by the injection of energy into the domain
due to the (perturbed) body stresses.  Instead we interpret
the instability as a consequence of the velocity
profiles set up by the conditions.  In other words, since
$\dot  E_{{\rm visc}}$ is positive definite, this
process experiences growth entirely due to the
extraction of energy from the shear as embodied by
the domain integral term $\dot E_{{\rm shear}}$.
The character of the eigenfunctions are seen
by inspecting the corresponding
profiles for small values of $\epsilon$
in Figure \ref{bc_comparison_plot}.  In both the
rigid and stress-free cases there are prominent boundary
layers appearing on the starside for both
the pressure and potential vorticity.  The critical
layer in the potential vorticity also appears here (see the inset
in Figure \ref{bc_comparison_plot}c and \ref{bc_comparison_plot}d) but is dwarfed
by the starside boundary layer.

\subsubsection{Zero PV-anomaly}\label{zero_Q}
  {This boundary condition may be envisioned as the starside boundary
counteracting any tendency for the development of any PV-anomaly
there.  Although this is somewhat artificial, we present here the
results of this investigation because these boundary conditions give solutions
that closely resemble those obtained for the calculation on the semi-infinite
domain calculation. }
As in the semi-infinite domain calculation,
instability occurs
when $F_\epsilon > 1 + 2q/\beta^2$,
and it scales as $\epsilon$ for
$\epsilon \ll 1$.
Inspection of the eigenfunctions in Figure \ref{bc_comparison_plot}
for $\epsilon \ll 1$,
especially the profiles for $\hat Q$, shows that (i) the perturbed potential
vorticity
is strongly localized in the critical
layers occurring where the real wavespeed approximately
equals the background flow speed, (ii) an additional boundary layer appears at
the starside boundary scaling like $\epsilon$ and, (iii) the
vorticity in the critical layer follows the
$\epsilon^{1/3}$ scaling determined
in the semi-infinite domain calculation.
We depict in Fig. \ref{growthrates_freaks}a a contour plot of
growth rates as a function of both the Reynolds number and
the inverse of the Froude number, $F_\epsilon$ for fixed values
of $\beta$ and  $k$.  The vertical axis may be understood
as representing a positive increase in the wave's speed
(see Eq. \ref{PV_wavespeed}).  For the parameters depicted
in Fig. \ref{growthrates_freaks}a ($\beta = 2, q= 3/2$), instability sets in for
$F_\epsilon >\sqrt{7/4}$.

\subsubsection{Zero radial PV-anomaly gradient}
\begin{figure}
\begin{center}
\leavevmode \epsfysize=13.cm
\epsfbox{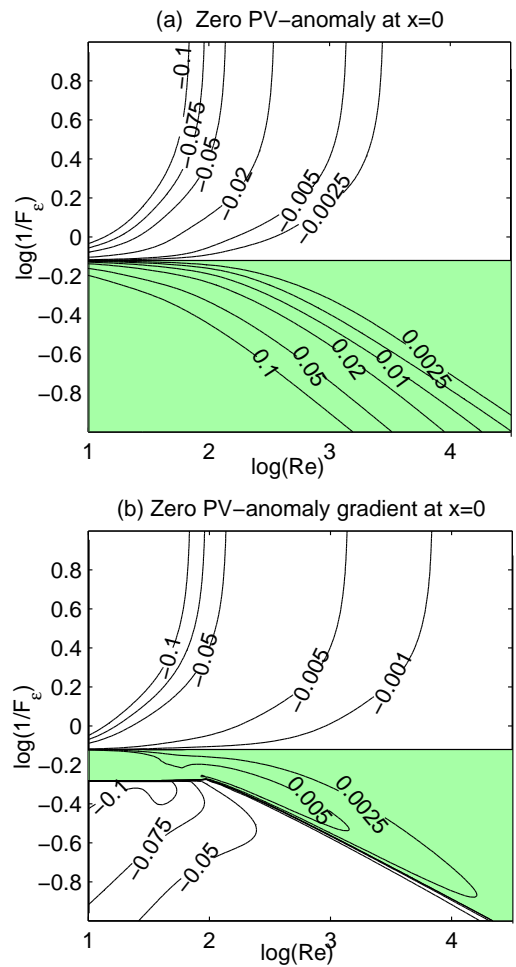}
\end{center}
\caption{{Same as Fig. \ref{growthrates_rigidstressfree} except:
(a) Zero PV-anomaly at $x=0$, (b) Zero PV-anomaly gradient at $x=0$.
}}
\label{growthrates_freaks}
\end{figure}

Requiring no radial gradient of the PV-anomaly on the starside boundary
is arguably the least physically realistic but we include it here, as in
the previous section, because it best
 reproduces
the asymptotic result of  the semi-infinite domain calculation.
Like in the previous section, where $\hat Q$ is set to zero there,
there is instability when $F_\epsilon^2 > 1$, a critical
layer emerges which also scales as $\epsilon^{2/3}$ for small
values of $\epsilon$.  However, the boundary layer appearing
near the starside boundary for the calculation
of Section \ref{zero_Q} vanishes here.  Finally, for small
values of $\epsilon$ (i.e. Re$^{-1}$) the Reynolds Orr
expression for these disturbances takes on the same leading
form as (\ref{Reynolds_Orr_semiinfinite})
in Section \ref{normal_modes_semi_infinite}; and this includes
the character of $\dot E_{{\rm stress}}$.
We note that growth
rates here are nearly identical to the growth rates
determined in Section \ref{zero_Q}.  Figure \ref{growthrates_freaks}b
shows the landscape of instability (for $\beta=2,q=3/2$)
and we see that instability also sets in when $F_\epsilon > 1 + 2q/\beta^2$
(i.e. here for $F_\epsilon >\sqrt{7/4}$)
but that it is bounded above by a more complicated function of
Re.
We note that the critical layer becomes harder to distinguish
as viscosity (that is, Re$^{-1}$ or $\epsilon$) is made larger.



\section{Discussion and Reflections}

\subsection{On the Tollmien Schlichting wave analogy}

  {T-S waves appear in flows that are are wall-bounded at least on one
boundary.  In the classic analysis done for Blasius boundary layers
(e.g. Schlicting, 1968) the instability is a
solution of the 2D Orr-Sommerfeld equation.
The mechanics leading to instability
is understood to arise
from the action of a purely viscous mode interacting with a nearly
inviscid mode (Baines, Majumdar \& Mitsudera, 1996).  The global
velocity profiles set up are such that the relative phase between
the horizontal and (plate) normal-velocities promotes extraction of
energy from the shear which then leads to energetic growth (i.e. $\dot E_{{\rm shear}} > 0$).
The 3D instability studied here
shares some major similarities to classic T-S waves: (i) the equation
governing the dynamics of the potential vorticity modes (\ref{the_equation_of_motion})
has similar structure to the 2D Orr-Sommerfeld equation, (ii) the
instability emerges for both no-slip and free-slip boundary conditions
(but not limited to these).}
  {What stands out in our minds is that, although
classical unstable T-S waves come about in a wide variety of
background flows excluding plane-Couette flow
(Baines, Majumdar \& Mitsudera, 1996, Schmid \& Henningson, 2000), the
instability here is present for a plane-Couette type of flow profile.}
\par
  {The results here compare
qualitatively to the results of rotating Blasius boundary layers.  For
example, Yecko \& Rossi (2004) show that three dimensional modal instability preferentially
emerges in anticylonic rotating Blasius flow when the vertical wavenumber
of the disturbance is large in comparison to its streamwise (azimuthal)
wavenumber (e.g. see Figure 8b in Yecko \& Rossi, 2004). By comparison,
the asymptotic scalings we have implemented spotlights
dynamics characterized by these same spatial scale disparities.
Thus despite the differences in the problems investigated between
these two studies (i.e.
the inclusion of gravity and entropy gradients and the differing
base velocity profiles)  the similarities in
the circumstances for instability onset suggest that
such processes may be more general in environments like disks - especially
near the disk-star boundary.
}

\subsection{On the assumed constancy of $g$ and $N$}
  {
To make the analysis we have exposed here analytically
possible we assumed that the vertical component of gravity
and the Brunt-V\"{a}is\"{a}l\"{a} to be constant with
respect to the disk vertical coordinate, (\ref{g_assumption}) which
permits us to assume separable normal-mode solutions.
On the other hand, real disks (including their small sections) are characterized
by vertically varying values of $g$ and $N$ and this
means that, in general, one cannot assume separable solutions in $z$ and $x$, in particular.
\footnote{In other words by assuming $g$ and $N$ to be constant
we are able to assume solution form $\Pi = X(x)Z(z)T(t)Y(y)$
while if $g$ and $N$ are z-dependent one can (at best) assume
a solution in the form $\Pi = \tilde\Pi(x,z) T(t) Y(y)$
where $\tilde\Pi(x,z)$ is the non-separable structure function.}
 We have  checked
that the results obtained in the limit where the viscosity
parameter is small (i.e. $\epsilon \ll 1$) still holds when $g$ and
$N$ are taken to be correctly $z$ dependent.  Accordingly we have repeated
the asymptotic calculation described in Section 4.2 where, in addition,
we restricted our attention to finite vertical domains by imposing either
velocity or pressure conditions on the vertical boundaries.  Such
disturbances will be characterized by vertical overtones
labeled by an overtone wavelength $\beta_n$ - which should be
thought of as being analagous to the vertical wavenumber $\beta$
we assumed in Section 4.
The asymptotic
calculation shows that instability  sets in so long as a
vertically weighted Froude number, $\tilde F_\epsilon(\beta_n)$,
approximately exceeds one.
The calculation is far more lengthly and does not add
any new qualitative details to the one presented in this work and it
is for this reason we have omitted it from the current exposition and we will
expand upon it in a future study.}

\subsection{Relationship to the viscous pulsational instability}
  {In an axisymmetric study of an shearing sheet section
of an accretion disk of constant temperature,
Latter \& Ogilvie (2006) argue that the VPI (Kato, 1978)
is most likely to manifest itself
through the destabilization of an even structured f-mode.
The disturbances
become unstable because the viscous perturbations transfer energy
from the shear into the acoustic mode through the perturbed stress  $T_{xy}'$.
Because f-modes
are characteristically inertial-acoustic waves,
they are the likely candidates for this instability
since their vertical structures are the simplest which, in turn,
result in minimizing dissipative losses.
We observe that $T_{xy}'$ is proportional to the pressure fluctuation.
This fluctuating stress affects the evolution of the horizontal velocity
perturbation by extracting/adding energy into the disturbance.
As the horizontal velocity disturbances are not in general in phase
with the pressure fluctuations, especially for inertial/acoustic modes,
the possibility for overstability is manifest.
However, the PV-anomalies examined here are
distinct from inertial-gravity and inertial-acoustic modes (e.g.
Ogilvie, 1998, Tevzadze et al., 2004).  The horizontal velocity
perturbations of PV anomalies are proportional to the radial
gradient of the perturbation pressure.  Since the latter quantity
is proportional to a decaying exponential (i.e. $\sim e^{-k_f x}$),
it follows that the phase between $T_{xy}'$
and the energy in the PV-anomaly will be $\pi$ radians
out of phase with each other (see Section 4.1). Thus the fluctuating
stress $T_{xy}'$ behaves to stabilize a PV-anomaly.
resulting in a stabilizing relationship as we observe in Section 4.1.
As such we understand the destabilization of the PV-anomaly as
as being distinct from the instability leading to the VPI.

\subsection{Summary and Implications}
  {For cold disk systems, i.e.
those in which magnetic effects are not active, the prospects
of identifying instability mechanisms appear to be far from exhausted.
We have tried to argue that certain previously considered non-magnetic
instabilities need not be discarded as candidate mechanisms driving activity
for disks.  Indeed the SRI and PPI instabilities, which emerge as
the interaction of edgewaves along cylinder/channel walls, could in principle
operate in real disks so long as there exists, \emph{in general}, interacting waves
propagating separated from each other by a wave-evanescent region.
We have demonstrated here
another possible mechanism - that the existence of disk analogs
of unstable Tollmien-Schlichting
waves could also manifest themselves in real disk systems.  We have
carried out the calculation within a model shearing box 
in which we have imposed a single
boundary on one side.  True disks have boundary layers separating
stars from the disk which are probably far more complicated
(Regev \& Bertout, 1995) than the model we have presented here.
Nonetheless, far from being a proof, we have demonstrated
in this asymptotically simplified model
that such a dynamical processes is, at least, feasible.
It is no stretch of the imagination
to suppose that analogous unstable waves may exist
near the vicinity of the star-disk boundary layer.
We add a final reflection.  Classical unstable T-S waves emerge
in flows with compliant boundaries showing that
such instabilities are robust and persist even if the boundaries
have a certain amount of elasticity to them (Carpenter \& Garrad, 1985) -
although compliant walls delay the onset of instability to higher
Reynolds numbers.  As a star-disk boundary is probably not a rigid body
transition, it would be beneficial to investigate and/or
model these disk analog T-S waves by considering starside boundary
conditions that are appropriately compliant as well.}
\par
  {The perspective we have adopted therefore can be broken down
into two parts.  The first is that (as in the recent studies
of Kleiber \& Glatzel 1999, and Latter \& Ogilvie 2006)
we have expanded the exploration of the possible
destabilizing role viscosity can play.  Viscosity does not always have to
stabilize disturbances as there are velocity profiles, dictated by
boundary conditions, wherein destabilization occurs counter to one's usual
physical intuition.
The
second is that experimentation with boundary conditions, even
within the context of the shearing box environment, followed
by concerted effort toward understanding and clarifying their
effects is a worthwhile endeavor given our lack of complete
knowledge about the boundaries of real disk systems (a situation
which is strongly contrasted by what is encountered in laboratory/terrestrial
flows).}\par
  {If turbulent stresses in
cold disks are driven by the MRI resulting
in effective values of $\alpha\sim 10^{-3} - 10^{-4}$
(Ogilvie, 2003, King, Pringle \& Livio, 2007),
then T-S waves like the sort here could emerge
as a secondary instability.  This possibility is made
manifest because the T-S wave instability grows faster
in proportion to the value of $\alpha$ until about
 a value of $0.1$.
\par

\section{Acknowledgements}
The authors are indebted to the valuable comments and suggestions of
the anonymous referee.
The authors would like to thank the Israeli Science Foundation for
making this research possible.
OMU also acknowledges that this research was partly supported by
BSF grant 2004087 and ISF grant 1084/06.
OMU also
thanks the Dead Sea Regional Council and the Ein Gedi Kibbutz for
their hospitality and Phil Yecko for suggestive conversations.

\appendix

\section{Scaling arguments leading to the viscous large shearing box equations}
The derivation of the VLSB equations follows the procedure
executed in Umurhan \& Regev (2004). The dimensional equations of
motion in cylindrical coordinates in a frame of constant rotation
are, in component form, given by the following \beqa
\frac{du_{_r}}{dt} -\tilde\Omega_0^2 r - 2\tilde\Omega_0\tilde
u_{_\phi} -\frac{\tilde u_{_\phi}^2}{r} &=&
-\frac{1}{\rho}\frac{\partial P}{\partial r} -\frac{\partial
\Psi}{\partial r}
+N_{_r}, \label{radial_eqn}\\
\frac{d\tilde u_{_\phi}}{dt} + 2\tilde\Omega_0 u_{_r}
+\frac{\tilde u_{_\phi} u_{_r}}{r} &=& -\frac{1}{\rho
r}\frac{\partial P}{\partial \phi} -\frac{1}{r}\frac{\partial
\Psi}{\partial \phi}
+N_{_\phi}, \label{azimuthal_eqn}\\
\frac{du_{_z}}{dt}  &=& -\frac{1}{\rho}\frac{\partial P}{\partial
z} -\frac{\partial \Psi}{\partial z} +N_{_z}\label{vertical_eqn},
\eeqa the equations of mass continuity and entropy \beqa
\frac{d\rho}{dt} + \rho \left(\frac{1}{r}\frac{\partial r
u_{_r}}{\partial r} +\frac{1}{r}\frac{\partial \tilde
u_{_\phi}}{\partial \phi}
+ \frac{\partial u_{_z}}{\partial z}\right) &=& 0, \\
\frac{d \Sigma}{dt} &=& Q, \eeqa with the operator definition
\[
\frac{d}{dt} \equiv  \frac{\partial}{\partial t} +
u_{_r}\frac{\partial}{\partial r} + \frac{\tilde
u_{_\phi}}{r}\frac{\partial}{\partial \phi} +
u_{_z}\frac{\partial}{\partial z}.
\]
$u_{_r},\tilde u_{_\phi},u_{_z}$ are the radial ($r$), azimuthal
$(\phi$) and vertical ($z$) velocities as observed in the
\emph{rotating} frame.  The rotation rate $\tilde\Omega_0$ is set
to the rotation at some fiducial radius $r=R_{_0}$. The curious
notation on $\tilde u_{_\phi}$ is meant to indicate that the
azimuthal velocity observed in the laboratory frame (denoted by
$u_{_\phi}$) would be related to its velocity in the rotating
frame by $u_{_\phi} = \tilde\Omega_0 r + \tilde u_{_\phi}$. The
entropy is defined to be $\Sigma \equiv C_{_V}\ln P\rho^{-\gamma}$
in which $C_{_V}$ is the specific heat at constant volume,
$\gamma$ is the ratio of specific heats, that is $\gamma \equiv
C_{_V}/C_{_P}$ in which $C_{_P}$ is the specific heat at constant
pressure. $Q$ is a heat function representing some sort of
non-adiabatic processes that are relevant to the disk. \par The
gravitational potential $\Psi$ is written in the following unusual
form in order to effect some generality when it comes to the shear
it induces upon the steady state flow,
\beq \Psi =
\frac{\tilde\Omega_{_0}^2 R_{_0}^2}{\left( \frac{r^2}{R_{_0}^2}
+\frac{z^2}{R_{_0}^2}\right)^{q-1}}.\label{potential_definition}
\eeq
For the realistic Keplerian flow profile $q = 3/2$ and
$\tilde\Omega_{_0}^2 = GM_{*}/R_{_0}^3$, where $M_{_*}$ is the
mass of the central object.  However, $\tilde\Omega_{_0}$ is in
general a function of $q$ and only when $q=3/2$ it is to be
interpreted as the "Keplerian" rotation rate.
\par
The viscous moments are \beqa \rho N_{_r} &=&
\frac{1}{r}\frac{\partial r t_{_{rr}}}{\partial r} +
\frac{1}{r}\frac{\partial r t_{_{r\phi}}}{\partial \phi}
-\frac{t_{_{\phi\phi}}}{r} +\frac{\partial t_{_{rz}}}{\partial z}, \\
\rho N_{_\phi} &=& \frac{1}{r^2}\frac{\partial r^2
t_{_{r\phi}}}{\partial r} + \frac{1}{r}\frac{\partial
t_{_{\phi\phi}}}{\partial \phi}
+\frac{\partial  t_{_{z\phi}}}{\partial z}, \\
\rho N_{_z} &=& \frac{1}{r}\frac{\partial r t_{_{rz}}}{\partial r}
+ \frac{1}{r}\frac{\partial  t_{_{\phi z }}}{\partial \phi}
+\frac{\partial  t_{_{zz}}}{\partial z}, \eeqa
along with the
viscous stresses $t_{_{ij}}$,
\beqa t_{_{rr}} &=& 2\eta
\frac{\partial u_{_r}}{\partial r}
+\left(\zeta - \frac{2}{3}\eta\right)\nabla_{r}\cdot{\bf u}, \\
t_{_{\phi r}} = t_{_{r\phi}} &=& \eta\left[ \frac{1}{
r}\frac{\partial}{\partial r}\left( \frac{\tilde u_{_\phi} +
r\tilde\Omega_{_0}}{r}\right) +\frac{1}{r}\frac{\partial
u_{_r}}{\partial \phi}
\right], \\
t_{_{z r}} = t_{_{rz}} &=& \eta\left( \frac{\partial
u_{_z}}{\partial r} +\frac{\partial u_{_r}}{\partial z}
\right), \\
t_{_{\phi\phi}} &=& 2\eta \left(\frac{1}{r} \frac{\partial \tilde
u_{_\phi}}{\partial \phi} +\frac{u_{_r}}{r}\right)
+\left(\zeta - \frac{2}{3}\eta\right)\nabla_{r}\cdot{\bf u}, \\
t_{_{z \phi}} = t_{_{\phi z}} &=& \eta\left( \frac{\partial \tilde
u_{_\phi}}{\partial z} +\frac{1}{r}\frac{\partial u_{_z}}{\partial
\phi}
\right), \\
t_{_{zz}} &=& 2\eta \frac{\partial u_{_z}}{\partial z}
+\left(\zeta - \frac{2}{3}\eta\right)\nabla_{r}\cdot{\bf u}, \eeqa
in which
\[
 \nabla_{r}\cdot{\bf u} \equiv
 \frac{1}{r}\frac{\partial r u_{_r}}{\partial r}
 +\frac{1}{r}\frac{\partial \tilde u_{_\phi}}{\partial \phi}
 +\frac{\partial r u_{_z}}{\partial z}.
\]
Because the turbulent viscosity within a disk is presumed to be
driven by some sort of shear process (either the MRI or a
subcritical hydrodynamic transition) the bulk viscosity $\zeta$
will be set to zero hereinafter. According to the classic proposal
of Shakura \& Sunyaev (1973), the shear viscosity is parametrized
as \beq \eta = \frac{2}{3}\frac{P}{\Omega}\alpha \eeq where
$\Omega = \tilde\Omega_0(R_0/r)^q$ is the steady-rotation rate induced by the generalized
potential (\ref{potential_definition}).  When we consider
Keplerian flows we write $\Omega_{_K} \equiv \Omega(q=3/2)$ (and
see below).  The parameter $\alpha$ is a tunable order 1 quantity.
\par
We proceed from here onto non-dimensionalization.  We shall
consider dynamics as taking place in a small region around the
point $r=R_{_0}, \phi = \phi_{_0}, z=0$, which we shall refer to
as \emph{the box}.  We let $\varepsilon$ measure the non-dimensional
size of this box, i.e.,
\beqa
& & R_{_0}\left(1-\frac{\varepsilon}{2}\right) \le r \le
R_{_0}\left(1+\frac{\varepsilon}{2}\right), \nonumber \\
& &  \phi_{_0} -
\frac{\varepsilon}{2} \le \phi \le \phi_{_0} + \frac{\varepsilon}{2}, \nonumber \\
& &  - R_{_0}\frac{\varepsilon}{2} \le z \le  R_{_0}
\frac{\epsilon}{2}. \nonumber
\eeqa
This motivates us to scale all the directions by $R_{_0}$ and to
define the nondimensional coordinates \beq \varepsilon x \equiv
\frac{r-R_{_0}}{R_{_0}}; \qquad \varepsilon y \equiv
(\phi-\phi_{_0});\qquad \varepsilon z \equiv \frac{z}{R_{_0}} . \eeq
where it is understood that $x,y,z$ are now taken to be order 1
quantities.
\par
Furthermore we suppose that all density quantities are scaled by
the reference density $\bar\rho$, pressure quantities are supposed
similarly scaled by $\bar\rho \tilde c_{_S}^2$ where $\tilde
c_{_S}$ is the dimensional scale of the sound speed of the box. In
parallel with this speed is the local rotation speed of the box
around the central object, $V_{_0} \equiv \tilde\Omega_{_0}
R_{_0}$: when $q=3/2$ this speed is sometimes referred to as
\emph{the local Keplerian speed of the disk}.  The fundamental
ansatz of cold thin disk theory is that the ratio of these two
quantities is small.  We, in fact, identify
\beq \varepsilon \equiv
\frac{c_{_S}}{V_{_0}}, \eeq
which is the classic parameter
measuring the thinness or "coldness" of the disk (Shakura \&
Sunyaev, 1973 and see recently Umurhan et al., 2006, for recent
exploitations of this parameter). {This, in turn, is a measure of
the disk's vertical scaleheight $H\equiv\varepsilon R_{_0}$.}
 We also
note here that by equating the size of the box to the "coldness"
of the disk means we are here looking at the viscous analog of the
so-called ``Large Shearing Box" equations formally developed in
Umurhan \& Regev (2004).
\par
Thus we propose that all velocities observed in this rotating
frame are scaled by the soundspeed, meaning to say that
\[
u_{_r} = c_{_S} u', \qquad \tilde u_{_\phi} =c_{_S} \tilde v',
\qquad u_{_z} = c_{_S} w',
\]
where $u',w',v'$ are order 1 nondimensionalizations of the
corresponding velocities.  \emph{Note that in these scalings we
are saying that all velocities observed in the rotating frame are
order $\epsilon$ smaller than the basic rotation speed of the box
$V_{_0}$.}  This seemingly obvious point is emphasized because the
magnitude scale of the accretion and meridional flow induced by
the turbulent viscosity is order $\varepsilon^2$ smaller than
$V_{_0}$ (Shakura \& Sunyaev, 1973, Kluzniak \& Kita, 1999). It
means that to the order to which matters are considered here, that
is as far as the generalized ``box" formalism is concerned (see
below), the effects of accretion and meridional flow are absent.
.\par Time and all advective derivatives are scaled
according to the local rotation time of the disk, i.e.
$\tilde\Omega_{_0}$. Before putting these scalings into the
governing equations we note that
\[
\tilde\Omega_0^2 r-\frac{\partial\Psi}{\partial r}
=2\tilde\Omega_0^2 R_{_0}\left[\varepsilon q x +
\order{\varepsilon^2}\right].
\]
Taking into account all of the nondimensionalizations, along with
writing $\rho \rightarrow \bar\rho \rho', P \rightarrow \bar\rho
c_{_S}^2 p'$ and understanding that $\rho',p'$ are the
nondimensionalized density and pressure quantities, we find that
the equations of motion (\ref{radial_eqn}-\ref{vertical_eqn})
become
\beqa & & \frac{\partial u'}{\partial t} +u'\frac{\partial
u'}{\partial x} +\tilde v'\frac{\partial u'}{\partial y}
+w'\frac{\partial u'}{\partial z} -2\tilde v' = \nonumber \\
& & \hskip 3.0cm 2 q x
-\frac{1}{\rho'}\frac{\partial p'}{\partial x} +N_{_r}' +
\order{\varepsilon}, \label{imt_radial}
\\
& & \frac{\partial \tilde v'}{\partial t} +u'\frac{\partial \tilde
w'}{\partial x} +\tilde v'\frac{\partial \tilde v'}{\partial y}
+w'\frac{\partial \tilde v'}{\partial z} + \ \ 2 u' = \nonumber \\
& &  \hskip 3.0cm -\frac{1}{\rho'}\frac{\partial p'}{\partial y}
+N_{_\phi}' + \order{\varepsilon},\\
& & \frac{\partial w'}{\partial t} +u'\frac{\partial w'}{\partial x}
+\tilde v'\frac{\partial w'}{\partial y} +w'\frac{\partial
w'}{\partial z}
 = \nonumber \\
& & \hskip 3.0cm  -\frac{1}{\rho'}\frac{\partial p'}{\partial z} - z
+N_{_z}' + \order{\varepsilon}. \eeqa
$N_{_r}',N_{_\phi}', N_{_z}'$
are as they appear in the text. To reiterate, the effects grouped
in the $\order{\epsilon}$ terms consist of curvature effects,
higher order corrections due to the central potential, and
turbulent viscosity induced accretion/meridional flow.  The
equation of continuity and entropy conservation also become
\beqa
& & \frac{\partial \rho'}{\partial t} +u'\frac{\partial
\rho'}{\partial x} +\tilde v'\frac{\partial \rho'}{\partial y}
+w'\frac{\partial \rho'}{\partial z} + \nonumber \\
& & \hskip 1.0cm \rho'\left( \frac{\partial
u'}{\partial x} +\frac{\partial \tilde v'}{\partial y}
+\frac{\partial w'}{\partial z}
\right) = 0 + \order{\varepsilon}, \\
& & \frac{\partial \Sigma}{\partial t} +u'\frac{\partial
\Sigma}{\partial x} +\tilde v'\frac{\partial \Sigma}{\partial y}
+w'\frac{\partial \Sigma}{\partial z} = 0 + \order{\varepsilon}.
\label{imt_entropy} \eeqa
To obtain the viscous large shearing box
equations presented in the text, i.e.
(\ref{lsb_continuity}-\ref{lsb_entropy}), we make the following
identifications and assumptions
\begin{itemize}
\item Drop all terms $\order{\varepsilon}$ and higher
from (\ref{imt_radial}-\ref{imt_entropy}).
\item Everywhere write $\tilde w' = -qx + w'$
in (\ref{imt_radial}-\ref{imt_entropy}) in order to eliminate the
sole $x$ term on the RHS of (\ref{imt_radial}).  The expression
$-qx$ is the background shear and will be felt by perturbations.
\item Write the density and pressures as being comprised of
a steady portion and a time dependent perturbed portion, i.e.
\beqa
 & & \rho' \rightarrow \rho_{_b}(x,z) + \rho(t,x,y,z), \nonumber \\
& & p' \rightarrow p_{_b}(x,z) + p(t,x,y,z). \nonumber
\eeqa
\item Specifically flag all Coriolis-like related effects with
the symbol $\Omega_{_0} \equiv 1$.
\end{itemize}

\section{Semi-infinite domain calculation details}\label{semi-infinite-details}
\subsection{Outer solution completion}\label{outer-solution-details}
As a result of this breakdown we write the solutions to this order according to which
side of $x_c$ one is on.  Formally then we say
\beq
\hat\Pi_1 = \hat\Pi_1^{(-)} = A_1^{(-)} e^{-k_{_F}x} + B_1^{(-)} e^{k_{_F}x} +
\hat\Pi_{1p}^{(-)} ,
\eeq
for $0\le x < x_c - \varepsilon_{-}$,
and
\beq
\hat\Pi_1 = \hat\Pi_1^{(+)} = A_1^{(+)} e^{-k_{_F}x} +
\hat\Pi_{1p}^{(+)} ,
\eeq
 for $x > x_c + \varepsilon_{+}$.
The particular solutions $\hat\Pi_{1p}^{(\pm)}$ satisfy
\beq
(\partial_x^2 - k_{_F}^2)\hat\Pi_{1p}^{(\pm)} = \frac{\Lambda A_0 e^{-k_{_F}x}}{i(x-c_0)}
\eeq
with solutions given by
\beqa
\hat\Pi_{1p}^{(-)} &=& \frac{A_0\Lambda}{2k_{_F}i}\biggl\{
e^{xk_{_F}-2c_{_0}k_{_F}}{\rm Ei}\left[2k_{_F}(c_{_0}-x)\right] \nonumber \\
& & \hskip 2.5cm -e^{-xk_{_F}}\ln\bigl(c_{_0}-x\bigr)\biggr \}, \\
\hat\Pi_{1p}^{(+)} &=& \frac{A_0\Lambda}{2k_{_F}i}\biggl\{
e^{xk_{_F}-2c_{_0}k_{_F}}{\rm Ei}\left[2k_{_F}(c_{_0}-x)\right] \nonumber \\
& & \hskip 2.5cm -e^{-xk_{_F}}\ln\bigl(x-c_{_0}\bigr)\biggr \},
\eeqa
where ${\rm{Ei}}(x)$ is the Exponential Integral function (Abramowitz \& Stegun, 1972).
We define
\[
\hat\Pi^{(-)}_{1p}\left(x_c^{-}\right) \equiv \lim_{x\rightarrow x_c^{-}}\hat\Pi^{(-)}_{1p},
\qquad
\hat\Pi^{(+)}_{1p}\left(x_c^{+}\right) \equiv \lim_{x\rightarrow x_c^{+}}\hat\Pi^{(+)}_{1p}.
\]
An analysis of these solution forms for $\hat\Pi^{(\pm)}_{1p}$ shows that
\[
\hat\Pi^{(-)}_{1p}\left(x_c^{-}\right) = \hat\Pi^{(+)}_{1p}\left(x_c^{+}\right).
\]
In formally writing these solutions
we purposely avoid the region described by
\[ x_c-\varepsilon_{_-} < x < x_c + \varepsilon_{+}, \]
which we will refer to as the {\em critical layer}.  The
size of this bounding region is such that
$\order{1} \gg \order{\varepsilon_{_\pm}} \gg \order{\epsilon}$.
It will be shown that the solutions on either side of this region will
asymptotically
match at lowest non-trivial order to the solution emerging from the critical layer itself.
Finally, the $u=0$ boundary condition at $x=0$ may be read from
(\ref{u_Pi_relationship}) after making use of the above relationships
\beqa
0 &=& -\sfrac{1}{2}c_0i\left[-k_{_F}A_1^{(-)} + k_{_F}B_1^{(-)} + \partial_x\hat\Pi^{(-)}_{1p}
|_{x=0}\right] \nonumber \\
& & \hskip -0.0cm +\sfrac{1}{2}ik_{_F}c_1A_0 - \sfrac{1}{2k_{_F}}\Lambda A_0
\nonumber \\
& &    - i\sfrac{1}{q}\left[A_1^{(-)}  + B_1^{(-)}
+ \hat\Pi^{(-)}_{1p}(0)\right]. \label{second_order_bc}
\eeqa
\par
\subsection{Critical layer calculation and matching}\label{critical-layer_calculation}
The calculation will be facilitated if we consider the evolution
of the residual potential vorticity quantity $\breve \Pi$ through
\beqa
\Pi &=& \breve \Pi +
A_0 e^{-2/q}\biggl[1 - \epsilon^{1/3}k_{_F}\xi +  \epsilon^{2/3} \sfrac{1}{2}k_{_F}^2\xi^2
\nonumber \\
& & \hskip 2.5cm -  \epsilon^{1}\sfrac{1}{6}k_{_F}^3\xi^3 +
\epsilon^{4/3}\sfrac{1}{24}k_{_F}^2\xi^4\biggr]. \nonumber
\eeqa
The polynomial terms in the above expression are the first five terms
of the series expansion of the lowest order outer solution, $\hat\Pi_0$
expressed in the vicinity of the critical layer.  We may now rewrite
(\ref{critical_layer_eqn}) instead in terms of $\breve\Pi$,
\beqa
\left(\frac{1}{i}\partial_\xi^4 - \xi\partial_\xi^2\right) \tilde \Pi &=&
   \epsilon^{2/3}\left[
\left(c_1\frac{\beta^2 + 2q}{i}\right)\partial_\xi^2 +
\xi k_{F}^2\right]\tilde\Pi  \nonumber \\
& & -\epsilon^{4/3}\frac{A_0e^{-2/q}\Lambda}{i} +
\order{\epsilon^{5/3}},
\eeqa
where $\Lambda$ is as defined in the text.
At $\order 1$ we have simply
\beq
{\cal L} \partial_\xi^2 \tilde \Pi_1 = 0, \qquad {\rm with} \qquad
{\cal L} \equiv \left(\frac{1}{i}\partial_\xi^2 - \xi\right).
\eeq
Homogeneous solutions of the operator ${\cal L} \partial_\xi^2$
of these involve integrals
of Airy Functions associated.  However, because these
are functions of complex arguments these two
solutions are rejected because they show exponential growth
as $\xi \rightarrow \pm\infty$ (e.g. Bender \& Orszag, 1978).
All homogeneous solutions involving these Airy functions are rejected henceforth.
 Thus we assume the following
solution expansion for $\breve\Pi$
\[
\tilde\Pi = \epsilon \tilde \Pi_1 + \epsilon^{4/3}\tilde \Pi_{4/3} + \cdots
\]
The order $\epsilon$ solution will be $\breve\Pi_1 = \tilde A_1$.  There is also
a solution at this order proportional to $\xi$ but it is rejected on account
of the fact that it will not match any corresponding solution from outside.
The equation
at order $\epsilon^{4/3}$ is
\beq
{\cal L}\tilde\Pi_{4/3} = -\sfrac{1}{i}\Lambda\tilde A_{0}.
\label{eqn_for_Pi43}
\eeq
with $\tilde A_0 = A_0 e^{-2/q}$.
With solution
\beq
\tilde\Pi_{4/3} = B_{4/3}\xi + \breve\Pi_{4/3}
\eeq
In which $\partial_\xi^2\breve\Pi_{4/3} \equiv \Xi$ and where
\beq
\Xi = -\Lambda \tilde A_0 \int_0^{\infty}{\exp{
\left(i\omega\xi-\sfrac{1}{3}\omega^3\right)}d\omega}.
\label{Xi_solution}
\eeq
Operating on $\Xi$ by ${\cal L}$
followed by an integration by parts verifies that $\breve\Pi_{4/3}$ is
the particular solution of (\ref{eqn_for_Pi43}).
We note immediately that $\Xi$ is bounded for all real values of $x$ by
$\Lambda \tilde A_0 \Gamma\left(\sfrac{4}{3}\right)$ since
\[
\int_0^{\infty}{\exp{
\left(i\omega\xi-\sfrac{1}{3}\omega^3\right)}d\omega}
\le
\int_0^{\infty}{\exp{
\left(-\sfrac{1}{3}\omega^3\right)}d\omega} = \Gamma\left(\sfrac{4}{3}\right).
\]
It is instructive to note the expression
\beq
\biggl[\partial_\xi \breve\Pi_{4/3}\biggr]^{\xi\rightarrow \Delta_{+}}_{\xi\rightarrow -\Delta_{-}}
= \int_{-\Delta_{-}}^{\Delta_{+}}\Xi d\xi,
\label{special_form_Xi}
\eeq
where  $\Delta_{\pm} > 0$ will be related to the expressions of $\varepsilon_{\pm}$
associated with the outer solutions (see next section).  This relationship
plays the role connecting the two homogeneous outer solutions.  When
$\Delta_{\pm} \gg 0$ an asymptotic evaluation of the integral shows that
\beq
\int_{-\Delta_{_-}}^{\Delta_{_+}}
{\Xi d\xi}
\sim
\frac{\Lambda}{i} A_0e^{-2/q}\left[\ln\Delta_{_+} - \ln\Delta{_-} + i\pi
\right].
\label{appx_Xi_integral}
\eeq
We note that there exists a phase factor proportional to $-i\pi$ that arises
from this operation.
This shall be explicitly referred to in the next section.
One final note: in the process of matching to the outer solutions the expression
$\epsilon^{4/3}\breve\Pi_{4/3}$ produces terms proportional to $\epsilon\ln\epsilon$ in
the outer region.  In order to have this appropriately matched (in this case, cancelled)
is the reason why there is an $\order{\epsilon\ln\epsilon}$ term in the critical layer
expansion (\ref{critical_layer_solution_expansion}).\par
We note that the PV-anomaly arising from this
critical layer appears at order $\epsilon^{2/3}$ when viewed in the
unstretched coordinate frame.  This is because the first non-trivial
contribution arising to the potential vorticity perturbation from this zone is
\beqa
\hat Q &=& (\partial_x^2 - k_F^2)(\tilde\Pi_0 + \cdots) \nonumber \\
&\sim & \epsilon^{4/3}\partial_x^2\breve{\Pi}_{4/3} + \cdots \nonumber \\
&=& \epsilon^{2/3}\partial_\xi^2\breve\Pi_{4/3} + \cdots \nonumber
\eeqa
Thus, while the size of the critical layer zone is order $\epsilon^{1/3}$
the magnitude of the PV-anomaly will be order $\epsilon^{2/3}$ the leading
order pressure perturbation.
\subsection{Matching}
We must reexpress the outer solution in a ``small" vicinity of the critical point $x_c$.
We consider first the solutions approaching from below, that is $x \rightarrow x_c^-$.
We have
\beqa
& & \hat\Pi\left(x\rightarrow x_c^{-}\right) = A_0 e^{-2/q}\biggl[1 - k_{_F}(x-x_c)
\nonumber \\
 & & \hskip 0.0cm
+\sfrac{1}{2}k_{_F}^2(x-x_c)^2
 - \sfrac{1}{6}k_{_F}^3(x-x_c)^3 + \sfrac{1}{24}k_{_F}^2(x-x_c)^4 + \cdots\biggr]
 \nonumber \\
& &  +  \epsilon\biggl[A_1^{(-)} e^{-2/q} + B_1^{(-)} e^{2/q} \nonumber \\
& & \hskip 1.0cm - k_{_F}\left(A_1^{(-)} e^{-2/q} - B_1^{(-)} e^{2/q}\right)(x-x_c) \biggr] \nonumber \\
& &  + \epsilon A_0 \frac{\Lambda}{ie^{2/q}}\left[(x-x_c)\ln(x_c-x)
+ (x_c-x)\right] \nonumber \\
& & + \epsilon\hat\Pi^{(-)}_{1p}\left(x_c^{-}\right),
\eeqa
while when approaching this point from above, that is as $x\rightarrow x_c^+$, we
find
\beqa
& & \hat\Pi\left(x\rightarrow x_c^{+}\right) = A_0 e^{-2/q}\biggl[1 - k_{_F}(x-x_c) + \sfrac{1}{2}k_{_F}^2(x-x_c)^2
\nonumber \\
& & \hskip 1.5cm  - \sfrac{1}{6}k_{_F}^3(x-x_c)^3 + \sfrac{1}{24}k_{_F}^2(x-x_c)^4 + \cdots\biggr]
 \nonumber \\
& &  +  e^{-2/q}\epsilon\biggl[A_1^{(+)} - k_{_F}A_1^{(+)}(x-x_c) \biggr] \nonumber \\
& & +\epsilon A_0 \Lambda \frac{1}{ie^{2/q}}\left[(x-x_c)\ln(x-x_c) - (x-x_c)\right] \nonumber \\
& & + \epsilon\hat\Pi^{(+)}_{1p}\left(x_c^{+}\right).
\eeqa
\par
Now we do the same to the critical layer solution.  We restore the inner coordinate $\xi$
in terms
of $x-x_c$ and take the limit of small $\epsilon$ (as is standard practice in boundary
layer theory, Bender \& Orszag, 1999) revealing
\beqa
 \hat\Pi &=& \hat A_0 + k_{_f}B_{1/3}(x-x_c) + \sfrac{1}{2}\hat A_0k_{_F}^2(x-x_c)^2
 \nonumber \\
 & &  + \sfrac{1}{6}B_{1/3} k_{_F}^2(x-x_c)^3
   + \sfrac{1}{24}\tilde A_{0} k_{_F}^4(x-x_c)^4 \nonumber \\
 & & + \epsilon \left[\hat A_1 + \hat B_{4/3}(x-x_c)\right] +
\epsilon^{4/3}\breve\Pi_{4/3}\left(\frac{x-x_c}{\epsilon^{1/3}}\right).
\eeqa
Identifications are made respecting powers of $\epsilon$ and $x-x_c$.
The $\order{1}$ matchings are straightforward since the solutions coming in from
the left and from the right of $x_c$ are the same:
\beq
\hat A_0 = A_0 e^{-2/q},\quad \hat B_{1/3} = -e^{-2/q} A_0, \label{matching_1}
\eeq
while the remaining two terms, proportional to $(x-x_c)^3$ and $(x-x_c)^4$
are satisfied given the above assignments in (\ref{matching_1}).  At $\order{\epsilon}$
we find first that
\beqa
& & e^{-2/q}A_1^{(-)} + e^{2/q}B_1^{(-)} + \hat\Pi^{(-)}_{1p}(x_c^{-}) = \hat A_1 =
\nonumber \\
& & \hskip 2.5cm e^{-2/q}A_1^{(+)}
 + \hat\Pi^{(+)}_{1p}(x_c^{+}), \nonumber
\eeqa
because $\hat\Pi^{(-)}_{1p}\left(x_c^{-}\right)=\hat\Pi^{(+)}_{1p}\left(x_c^{+}\right)$
the above relationship implies
\beq
A_1^{(-)} + B_1^{(-)}e^{4/q} = A_1^{(+)}.
\label{first_condition}
\eeq
To complete the matching we must prepare the final term
$\epsilon^{4/3}\breve\Pi_{4/3}$.  It is asymptotically
correct to do
this by considering the derivative of these terms up to the bounding region of
of the critical layer, i.e. $1 \gg |x-x_c| \gg 0$.  Thus we consider the derivatives
as one approaches this zone (and measured by the coordinates $\varepsilon_{\pm}$) as
\beqa
& & \epsilon\biggl[\partial_x \hat \Pi_1\biggr]_{x\rightarrow x_c - \varepsilon_{-}}
=\epsilon A_0\frac{\Lambda}{ie^{2/q}}\ln\varepsilon_{-} - \nonumber \\
& & \hskip 2.5cm -\epsilon k_{_F}\left(A_1^{(-)}e^{-2/q}
-B_1^{(-)}e^{2/q}\right),
\eeqa
that is, approaching $x_c - \varepsilon_{_-}$ from below
and
\beq
\epsilon\biggl[\partial_x \hat \Pi_1\biggr]_{x\rightarrow x_c + \varepsilon_{+}}
=\epsilon A_0\frac{\Lambda}{ie^{2/q}}\ln\varepsilon_{+} - \epsilon  k_{_F}A_1^{(+)}e^{-2/q},
\eeq
that is, approaching $x_c + \varepsilon_{_+}$ from above.
Subtracting the two expressions gives
\beqa
& & \epsilon\biggl[\partial_x \hat \Pi_1\biggr]^{x\rightarrow x_c + \varepsilon_{+}}_{x\rightarrow x_c - \varepsilon_{-}}
=\nonumber \\
& & - k_{_F}\epsilon\left[A_1^{(+)}e^{-2/q} -\left(A_1^{(-)}e^{-2/q} -
B_1^{(-)}e^{2/q}\right)\right] \nonumber \\
& & +
\epsilon A_0\frac{\Lambda}{ie^{2/q}}\left[\ln\varepsilon_{+}-\ln\varepsilon_{-}\right].
\label{outer_derivative_limit}
\eeqa
Now we must match this to the corresponding expression emerging from the interior
of the domain.  In other words we require similarly that
\beqa
& & \epsilon^{4/3}\biggl[\partial_x\breve\Pi_{4/3}\biggr]^{x\rightarrow x_c
 + \varepsilon_{+}}_{x\rightarrow x_c - \varepsilon_{-}}
 \rightarrow
 \epsilon\biggl[\partial_\xi\breve\Pi_{4/3}\biggr]
 ^{\xi\rightarrow\varepsilon_{+}/\epsilon^{1/3}}_{\xi\rightarrow- \varepsilon_{-}/\epsilon^{1/3}}
 \nonumber \\
& &  \hskip 2.0cm = \epsilon\int_{-\varepsilon_{-}/\epsilon^{1/3}}^{\varepsilon_+/\epsilon^{1/3}}
 \Xi d\xi,
 \label{inner_derivative_limit}
\eeqa
in which (i) the transition in orders of $\epsilon$ occurs because of the change of
variables from $x$ to $\xi$, (ii) we have identified $\Delta_{\pm} \leftrightarrow
\varepsilon_{\pm}/\epsilon^{1/3}$ and (iii) used (\ref{special_form_Xi}) in writing
the above expression.
We noted that the leading behavior of the integral for
$\varepsilon_{\pm}/\epsilon^{1/3} \gg 1$ (i.e. for the matching zone) is
\beq
\epsilon\int_{-\varepsilon_{_-}/\epsilon^{1/3}}^{\varepsilon_{_+}/\epsilon^{1/3}}
{\Xi d\xi}
\sim
\frac{\Lambda}{i} A_0e^{-2/q}\epsilon\left[\ln\varepsilon_{_+} - \ln\varepsilon_{_-} + i\pi
\right].
\label{appx_Xi_integral}
\eeq
Equating the RHS of (\ref{outer_derivative_limit}) and
(\ref{inner_derivative_limit}) and making use of the asymptotic form
(\ref{appx_Xi_integral})
we see that the offending logarithmic terms cancel leaving,
\beq
-k_{_F}\left[A_1^{(+)} -A_1^{(-)} + B_1^{(-)}e^{4/q}\right] =
A_0\frac{\Lambda}{i}\cdot i\pi = A_0\Lambda\pi. \label{second_condition}
\eeq
The meaning of this relationship is clear - the presence of the singular layer in the
flow means that the homogeneous outer region solutions must show a jump in their
derivatives (in proportion to the RHS of the above expression).  Another way
to interpret this is to recognize that this jump corresponds to the presence of
a vortex sheet at $x = x_c$.
\par
The complex wavespeed may be now obtained from (\ref{first_condition}), (\ref{second_condition})
and the boundary condition (\ref{second_order_bc}).  We note that for there
to be a non-trivial solution to (\ref{first_condition}) and (\ref{second_condition}) the
following
relationship,
\beq
-B_1^{(-)}e^{-4/q} = B_1^{(-)}e^{-4/q} + \frac{A_0\Lambda\pi}{k_{_F}},
\label{solvability_condition}
\eeq
must be satisfied.  This is the solvability condition.
The second matter we note is that
\beqa
\hat\Pi_{1p}^{(-)}(0) &=&\frac{A_0\Lambda i}{2k_{_F}}\left[\ln\left(\frac{2}{qk_{_F}}\right)
-e^{4/q}{\rm Ei}\left(\frac{4}{q}\right)\right], \nonumber \\
\partial_x\hat\Pi_{1p}^{(-)}\Bigr|_{x\rightarrow 0} &=&-\frac{A_0\Lambda i}{2}\left[\ln\left(\frac{2}{qk_{_F}}\right)
+e^{4/q}{\rm Ei}\left(\frac{4}{q}\right)\right]. \nonumber
\eeqa
Thus, the solvability condition (\ref{solvability_condition}) together with the above
expressions used in (\ref{second_order_bc}) combine to give
\beq
c_1 = i\frac{\Lambda}{qk_{_F}^2}\left[q+2e^{-4/q}{\rm Ei}\left(\frac{4}{q}\right)
-2i\pi e^{4/q}\right]. \label{complex_wavespeed}
\eeq
The form quoted in the text follows from restoring the definition of $\Lambda$ into the
above expression.


\begin{thebibliography}{}
\bibitem[2004]{AMN04}
Afshordi, N., Mukhopadhyay, B.,
\& Narayan, R.,
2005, ApJ, 629, 373
\bibitem[1996]{baines96}
Baines, P. G., Majumdar, S., \& Mitsudera, H., 1996, J. Fluid Mech. 312, 107
\bibitem[1994]{BM94}
        Baines, P. G. , \& Mitsudera, H., 1994,
        J. Fluid Mech. 276, 327
\bibitem[1999]{balmforth99}
Balmforth, N.J., J. Fluid Mech., 387, 97
 \bibitem[1999]{neil01}
  Balmforth, N.J. \& Piccolo, C., 2001, J. Fluid Mech., 449, 85
  \bibitem[1999]{neil01}
  Balmforth, N.J., Piccolo, C., \& Umurhan, O.M., 2001, J. Fluid Mech., 449, 115
 \bibitem[2003]{balbus_araa2003}
 Balbus, S.A., 2003, ARAA, 41, 555
 \bibitem[2005]{bmu00}
 Barranco, J.A., \& Marcus, P.S., 2005, ApJ, 623, 1157
\bibitem[1988]{bayly_review}
Bayly, B. J., Orszag, S. A., \& Herbert, T., 1988, Ann. Rev. Fluid Mech., 20, 359

  \bibitem[2007]{bodoA}
    Bodo, G., Tevzadze, A., Chagelishvili, G., Mignone, A., Rossi, P.,
    \& Ferrari, A. 2007, A\&A, 475, 51 (Bodo et al. A)

\bibitem[2007]{bodoB}
    Bodo, G., Chagelishvili, G., Murante, G., Tevzadze, A., Rossi, P.,
    \& Ferrari, A. 2007, {\tt arXiv:0705.3474v1[astro:ph]} (Bodo et al. B)

\bibitem[1985]{carpenter85}
Carpenter, P. W., \& Garrad, A. D. 1985, J. Fluid Mech., 155, 465
\bibitem[1960]{case}
Case, K.M., 1960, Phys. of Fluids, 3, 143
\bibitem[2003]{chag}
  Chagelishvili, G.D., Zahn, J.-P., Tevzadze, A. G., \& Lominadze, J.G., 2003,
  A\&A, 402, 401
\bibitem{charney_stern}
        Charney, J. G. \& Stern, M. E. 1962, J. Atmos. Sci., 19,159


\bibitem[1994]{bishop_davies}
        Davies, H.C., \& Bishop, C. H. 1994
        J. Atm. Sci., 51, 1930

\bibitem[doering]{doering}
Doering, C.R, Spiegel, E.A., \& Worthing, R.A., 2000, Phy. Fluids, 12, 1955
  \bibitem[1984]{drazin}
  Drazin, P.G. \& Reid, W.H., 1984, Hydrodynamic Stability, Cambridge.
\bibitem{dubrulle}
 Dubrulle B., Marie L. , Normand Ch., Richard D. ,  Hersant F.\&  Zahn  J.-P.
 2004, A\&A, 429, 1
\bibitem{FKR}
Frank J., King A.R. \& Raine D.J. 2002, Accretion Power in Astrophysics. Cambridge
Univ. Press, Cambridge



\bibitem[1965]{goldreich65}
        Goldreich, P., Lynden-Bell, D., 1965, MNRAS, 130, 125

\bibitem[1984]{ggn86}
        Goldreich, P., Goodman, J., \& Narayan, R. 1986, MNRAS, 221, 339

\bibitem[1987]{hayashi_young}
        Hayashi, Y.-Y. \& Young, W.R. 1987
        J. Fluid Mech.,  184, 477

\bibitem[1985]{hoskins}
        Hoskins, B. J., McIntyre, M. E., \& Robertson, A. W. 1985
        Quart. J. Roy. Meteor. Soc., 111, 877
 \bibitem[2001]{ik}
 Ioannaou, P.J. \& Kakouris, A., 2001, ApJ, 550, 931
 \bibitem[2006]{Ji}
    Ji, H., Burin, M., Schartman, E., \& Goodman, J. 2006
    Nature, 444, 343
\bibitem[2005]{gammie05}
Johnson, B.M. \& Gammie, C.F., 2005
ApJ, 635, 149
\bibitem{joseph}
Joseph, D.D., 2003, J. Fluid. Mech. ,479, 191
\bibitem[1978]{kato_78}
Kato, S., 1978, MNRAS, 185, 629
\bibitem[2004]{kersale1}
      Kersal\'e E., Hughes, D. W., Ogilvie, G. I., Tobias, S. M.,
      \& Weiss, N. O. 2004, ApJ, 602, 892
 \bibitem[2003]{klahr03}
 Klahr, H.H. \& Bodenheimer, P., 2003, ApJ, 582, 869
\bibitem[1999]{kleiber}
Kleiber, R. \& Glatzel, W., 1999, MNRAS, 303, 107
\bibitem[2007]{kpl}
      King, A. R., Pringle, J. E., \& Livio, M. 2007, MNRS, 376, 1740 (KPL07)

\bibitem{Kluzniak}
Klu\'zniak W. \& Kita, D. 2000, Three-dimensional structure of an
alpha accretion disk, {astro-ph/0006266}
\bibitem[2006]{latter_ogilvie} Latter, L. N., Ogilvie, G. I., MNRAS, 372, 1829
\bibitem[2005]{lesur}
Lesur, G. \& Longaretti, P-Y.,  2005, A\&A, 444, 25
\bibitem[2000]{li00}
Li, H., Finn, J. M., Lovelace, R. V. E. \& Colgate, S. A, 2000,
ApJ, 533, 1023
\bibitem{lions}
Lions, P.L. 1993, Limits incompressible et acoustique pour des fluides visqueux, compressible et isentropique, C.R. Acad. Sci. Paris Ser. I Math, 317,1197.
\bibitem[2007]{lithwick07A}
    Lithwick, Y. 2007
    ApJ, 670, 789 (Lithwick 2007 A)
\bibitem[2004]{muk04}
Mukhopadhyay, B., Afshordi, N., \& Narayan, R.,
2004, ApJ, 629, 383
\bibitem[1998]{ogilvie_98} Ogilvie, G. I., 1998, MNRAS, 297, 291
\bibitem[2001]{ogilvie_01} Ogilvie, G. I., 2001, MNRAS, 325, 231
\bibitem[2003]{ogilvie_proctor} Ogilvie, G. I., Proctor, M. R. E., 2003, J. Fluid Mech.,
476, 389
\bibitem[1984]{pp84} Papaloizou, J.C.B., \& Pringle, J.E. 1984, MNRAS, 208, 721



\bibitem[2007]{petersenA}
        Petersen, M. R., Julien, K., \& Stewart, G. R.  2007
        ApJ, 658, 1236 (Petersen et al. A)

\bibitem[2007]{petersenB}
        Petersen, M. R., Stewart, G. R., \& Julien, K.  2007
        ApJ, 658, 1252 (Petersen et al. B)

\bibitem{bertout95}
Regev, O. \& Bertout, C. 1995,
MNRAS,  272, 71-79
\bibitem[2001]{denis_thesis}
Richard, D., 2001, Thes\`e de doctorat, Universit\'e Paris 7.
\bibitem[1999]{richard_zahn_99}
Richard, D., Zahn, J.-P., 1999, A\& A, 347, 734
\bibitem[2001]{schlict}
Schlichting, H. \& Gersten, K. Boundary Layer Theory, 2001, Springer, p 94.
\bibitem[1989]{sakai}{}
        Sakai, S., 1989, J. Fluid Mech.  202, 149
 \bibitem[2001]{SP}
  Schmid, P. J. \& Henningson, D.S., 2000,
  Stability and Transition in Shear Flows, Springer
\bibitem{shalybkhov}
Shalybkov D. \&  R\"udiger G., 2005,  2005, A\&A, 438, 411
\bibitem{spiegel60}
Spiegel, E. A. \& Veronis G., 1960, ApJ, 131, 442
\bibitem{stewartson81}
Stewartson,  K., 1981, IMA J. Appl. Math., 27, 133
\bibitem[2003] {tevzadze} Tevzadze, A.G., Chagelishvili, G.D., Zahn, J.-P.,
Chanishvili, R.G., \&
Lominadze, J.G., 2003, A\&A, 407, 779
\bibitem[2005]{umurhan05} Umurhan, O.M. 2006, MNRAS, 365, 85
\bibitem[2006]{unrs06} Umurhan, O.M., Nemirovsky, A., Regev, O., \& Shaviv, G.,
2006, A\&A, 446, 1
\bibitem[2004]{UR04} Umurhan, O.M \& Regev, O. 2004, A\&A, 427, 855
\bibitem{yecko04}
Yecko, P.A. 2004, A\&A,  425, 385
\bibitem{yecko_rossi04}
Yecko, P.A. \& Rossi, M. 2004, Phys. Fluids,  16, 2322
\end{thebibliography}
\end{document}